\documentclass[12pt]{article}
\usepackage{amsmath}
\usepackage{times}
\usepackage{amssymb}
\usepackage{graphics}

\newcommand{\cross}{\times}

\renewcommand{\a}{\rho}
\renewcommand{\b}{\sigma}
\newcommand{\g}{\tau}

\newcommand{\R}{\ensuremath{{\mathbf R}}}
\newcommand{\A}{\ensuremath{{\mathbf A}}}

\newcommand{\E}{\ensuremath{{\mathcal E}}}
\newcommand{\F}{\ensuremath{{\mathcal F}}}

\newcommand{\J}{\ensuremath{{\mathcal J}}}

\newcommand{\HH}{\ensuremath{{\mathcal H}}}

\newcommand{\U}{\ensuremath{{\mathcal U}}}
\newcommand{\V}{\ensuremath{{\mathcal V}}}

\newcommand{\SU}{\ensuremath{{\mathcal S\mathcal U}}}
\newcommand{\C}{\ensuremath{{\mathbf C}}}
\newcommand{\OO}{\ensuremath{{\mathcal O}}}

\newcommand{\T}{\ensuremath{{\mathcal T}}}

\newcommand{\tr}{\ensuremath{{\mathrm{tr}}}}

\newcommand{\X}{\ensuremath{{\mathcal X}}}

\newcommand{\CR}{\ensuremath{{\mathcal C\mathcal R}}}

\begin{document}
\thispagestyle{empty}
\vspace*{30pt}\begin{center}
{\LARGE\textbf{A primordial theory}}\bigskip\vspace{10pt}\\
\textmd{George Sparling and Philip Tillman\\Laboratory of Axiomatics\\Department of
Mathematics\\ University of Pittsburgh\\Pittsburgh, Pennsylvania, USA\\\vspace{40pt}}
{\large\textbf{Abstract}}\\\vspace{20pt}
\begin{quote}\textbf{
\noindent We review the twistor approach to the Zhang-Hu theory of the four-dimensional Quantum Hall effect.  We point out the key role played by the group
$\textrm{Spin}(4,4)$, as the symmetry group of the boundary. It is argued that this group, which ignores the $\sqrt{-1}$ used in relativity and quantum
mechanics, is the focal point of a primordial theory, one where the Cartan concept of triality is paramount, from which the standard theories emerge via a
series of phase transitions of the Zhang-Hu fluid.  An important role will be played by the Jordan cross-product algebras, particularly the exceptional Jordan
algebra associated to the split octaves and by the associated Freudenthal phase space.  The geometry and Hamiltonian theory of these spaces is examined in
detail.  A possible link to the theory of massive particles is outlined.     }
\end{quote}
\end{center}

\thispagestyle{empty}\mbox{}\setcounter{page}{0}
\eject\noindent
\setcounter{section}{0}\setcounter{equation}{0}
\section*{Introduction} \subsection*{The theory of Zhang and Hu}The discovery of the theory of the Quantum Hall Effect in four dimensions by
Shou-Cheng Zhang and Jiang-Ping Hu has dramatically altered the landscape of modern physics \cite{Zh2}-\cite{Zh5}.  Their key result is that a fermionic $\SU(2)$
gauge fluid in four dimensions produces edge states, at the three-sphere boundary of the fluid, which behave as relativistic massless
particles of any and all helicities.  The implication is that the physics of our universe may be governed by such a fluid.  \\\\On studying
the work of Zhang and Hu, the first author realized that their formulas were strongly reminiscent of
basic formulas of twistor theory.  Indeed he was able to reformulate the theory entirely in twistor space, a complex projective
three-space \cite{Sp1, Pe1, Pe2}.  This manifold, of six real dimensions, is naturally a two-sphere bundle over the four-sphere employed by Zhang and
Hu.  In the twistor space, there is no gauge field and the fermions are single component entities, constituting an ordinary fermionic fluid. It
was shown that the boundary of the fluid in the twistor space is the standard $\CR$ hyperquadric of twistor theory: a manifold of five real dimensions, of topology
the product of a three-sphere and a two-sphere; it carries a Levi form of signature
$(+, -)$.  \\\\The hyperquadric is precisely the space whose space of interior complex projective lines (each of topology a
two-sphere) constitutes a four parameter set, which is naturally a compactified real Minkowski spacetime.  A three parameter subset of these lines
corresponding to a spacelike hypersurface foliates the hyperquadric, giving rise to the three-sphere boundary of the Zhang-Hu theory.  Then the edge
states correspond in twistor theory to $\CR$ sheaf cohomology classes: these in turn represent solutions of the zero-rest mass field
equations, whose helicity $s$ is governed by the homogeneity degree $n$ of the functions representing the sheaf sections, according to the
formula: $s = -\frac{1}{2}(n + 2)$.  Of particular interest is the sheaf cohomology appropriate to infinitesimal deformations of the
$\CR$-structure: these correspond to helicity minus two, or to gravity.  Thus the fundamental edge-states corresponding to deformations of the edge are
associated with gravity.  In this sense, it was argued that the Zhang-Hu theory is in essence, a theory of gravity. 
\subsection*{Philosophical and experimental issues}To extend the Zhang-Hu theory, one would like to see it generalized away from a conformally
flat background.  As indicated by the first author, one way to do this seems to be to develop the fermionic fluid in the context of null
cone hypersurface twistor spaces, since these possess analogues of the pseudo-Kahler hypersurface, which can serve as the boundary.  These spaces
(which as an ensemble form a manifold of ten real dimensions) have figured in the first author's proposal for linking superstring theory and twistor
theory \cite{Sp2}-\cite{Sp4}.  A difficulty arises with a potential loss of analyticity, which can squeeze the fluid towards the bounding hypersurface.  However
Zhang has indicated privately that such regions of non-analyticity could be modeled by global structures in the fluid, such as vortices.  Assuming that such
problems have been overcome, we are left with some fascinating and tantalizing philosophical issues:
\begin{itemize} \item The fermionic fluid is responsible for structure at the boundary, but is not itself at the boundary.  Spacetime \emph{arises} at
the boundary, but is not the boundary, instead being obtained by a projection from the boundary.  
\item It is probably  not legitimate to think of the fermionic fluid as in any way directly relativistic: relativity arises from excitations on 
the boundary that behave as massless particles.
\item In particular if the boundary disappeared either in the past or future, so would spacetime.
\item In a dynamical situation, it is possible that different fluid regions could merge or separate (analogous to the result of changing of the
electric or magnetic fields in the two-dimensional Quantum Hall Effect).  This would mean the merging or separation of "different universes".
\end{itemize}   
The theory tells us that we are in a situation like that in "Flatland" \cite{Ab1}: by our nature, we are creatures associated with the boundary and are not able
at the moment to experience the fermionic fluid directly.  But this does raise an important experimental question: 
\begin{itemize}\item  Can we invent some experiment which
will reveal the existence of the fluid to us?  Or in other words, is the twistor space in fact as "real" as spacetime itself?
\end{itemize} \eject\noindent 
There are several possible areas susceptible to experiment that spring to mind, where the fluid may make its influence felt:
\begin{itemize}\item The problem of non-local entanglement in quantum mechanics.
\item The problem of the origin and nature of mass.
\item  The problem of subtle asymmetries that exist, or may exist in nature: violation of $\mathcal{C}\mathcal{P}$, $\mathcal{T}$,
or even $\mathcal{C}\mathcal{P}\mathcal{T}$; the arrow of time; the apparently large excess of matter over anti-matter; the rather low density of magnetic
monopoles.
\end{itemize}
A useful warm-up exercise would be to return to the genesis of the Zhang-Hu theory, the two-dimensional Quantum Hall Effect.  Here we deal
with a physically realizable fermionic fluid made of electrons.  As shown by Mike Stone,  the boundary theory is a two-dimensional
relativistic string theory \cite{St1}.  Suppose we imagine ourselves to be entities governed by the boundary physics: could we conduct some experiment that would
convince us of the reality of the fermionic fluid? \\\\ Another question relevant particularly to the picture in curved spacetime is: \begin{itemize}
\item Is the fluid (or its boundary) in some sense \emph{cohomological}?
\end{itemize}  
To illustrate, consider the prototype of topological ideas in physics: the Dirac string \cite{Di1,Di2}.  If, following Dirac, we try to construct a vector
potential for the field of a magnetic monopole, then we are doomed to failure, unless we are prepared to allow for a string singularity
stretching from the monopole to infinity (or to some other oppositely charged monopole).  But the route taken by the string is
up to us to choose:  any two strings in the same homotopy class count as equivalent.  \\\\A similar (related) situation has occurred before in twistor
theory, where a complex space bifurcates, being non-Hausdorff at the bifurcation \cite{Pe5, Ma1}.  But the precise place of bifurcation is movable, by appropriate
analytic continuation.  The model here is a bleb on a car tire. This can be pushed down but never entirely eliminated.  So here it seems necessary to
require the boundary of the fluid to be analogous to the Dirac string.  Regarded one way it would give the pseudo-Kahler structure for one null cone
in spacetime.  Regarded another way it would give the null structure for another point, etc.  Then, in some sense, the fluid boundary would be
observer dependent.   The current work by physicists in the area of membranes seems to be connected here: the membranes seem to be taken
literally, yet they have a strong cohomological flavour, usually depending on the non-vanishing of some homotopy class for a stable "existence".
\eject\noindent 
\subsection*{The present work: the problem of $\sqrt{-1}$}
In the present work, we take as point of departure a detail of the twistor approach to the Zhang-Hu theory.  This is nature of the group that arises
governing the basic structure of the fluid boundary.  The $\CR$ hyperquadric of the theory can be given by the equation $Z^\alpha\overline{Z}_\alpha  =
0$, where $Z^\alpha$ is a vector in $\C^4$ and we are using abstract indices.  The conjugation $Z^\alpha \rightarrow \overline{Z}_\alpha$ is of type
$(2,2)$, so that the \emph{complex} linear invariance group of the hyperquadric is the sixteen-dimensional real Lie group $\U(2,2)$, which in turn maps
canonically to the conformal group of compactified Minkowski space-time.  It emerges that the natural way to write down amd analyze the $\CR$
structure, is in terms of an ensemble of vector fields:
\[ \overline{E}^{\alpha\beta} = Z^{[\alpha}\overline{\partial}^{\beta]}, \hspace{10pt}E_{\alpha\beta} =
\overline{Z}_{[\alpha}\partial_{\beta]},  \hspace{10pt} E_\beta^\alpha = Z^\alpha\partial_\beta - \overline{Z}_\beta \overline{\partial}^\alpha.\]
Here $\partial_\beta = \frac{\partial}{\partial Z^\beta}$ and $\overline{\partial}^\alpha = \frac{\partial}{\partial \overline{Z}_\alpha}$ and
brackets around tensor indices represent idempotent skew-symmetrization over those indices.  It is easy to check that each of these operators kills
the defining equation of the
$\CR$ hypersurface, so is tangent.  Then the anti-hermitian operators
$E_\beta^\alpha$ generate the expected $\U(2,2)$ symmetry of the hyperquadric.   However the $\CR$ structure itself is directly expressed by the
operator $\overline{E}^{\alpha\beta}$: the tangential Cauchy-Riemann equations on a function $f$ are just the equations $\overline{E}^{\alpha\beta}f =
0$.  Abstractly, the operator $\overline{E}^{\alpha\beta}$ has together with its conjugate 12 real components and it is easy to see
that $E_{\alpha\beta}$, $\overline{E}^{\alpha\beta} $ and $ E_\beta^\alpha $ together generate the twenty-eight dimensional real Lie
algebra
$\textrm{Spin}(4,4)$, which is the group of all \emph{real} linear transformations of twistor space, preserving the hyperquadric.  At this point, we recall
remarks of Sir Roger Penrose to the effect that the complex number $i = \sqrt{-1}$ occurs naturally in fundamental physics in two places \cite{Ha2}: 
\begin{itemize}\item In quantum mechanics: where, given the Hamiltonian $H$, the state vector $\psi$ evolves in time $t$ by $\psi\rightarrow
e^{iHt}\psi$. Also we need to be able to take
\emph{complex} linear combinations of states: $i\psi $ is defined for any state $\psi$.
\item In the theory of spacetime, where the space of rays of the null cone through a point inherits from the space-time conformal structure a natural
complex structure (\emph{only} in four dimensions).  This fact is exploited in relativity, particularly through the formalism of Ted Newman and
Penrose \cite{Ne1}.
\end{itemize}  
A common thread to these two occurrences of $\sqrt{-1}$ is the spinor, which can be used naturally to represent the state of a spinning particle and to
understand the geometry of a null vector.  Penrose has made the point that these two occurrences of $\sqrt{-1}$ are in some sense the same and that
this fact requires explanation. \eject\noindent Thus one might expect that in a theory combining quantum mechanics and gravity, complex numbers would
be important.  The multiplication of a twistor by $i$ is deeply related to both ideas, since, on the one hand, the twistor is expressed using
geometrical spinors and on the other habd, if the quantum states of particles are represented by functions of twistors, then multiplying the twistor by $i$,
will multiply the state by
$i^n$, where
$n$ is the (integral) homogeneity of the twistor function.  However, if we now contemplate the boundary structure of the Zhang-Hu theory, we are led
to an alternative working hypothesis:  
\begin{itemize}\item The reason why the two occurrences of $\sqrt{-1}$ are the "same" is that the $\sqrt{-1}$ arises in both areas, simultaneously, in
a transition  from a single theory.
\item This theory would not be a theory of quantum gravity per se.  Instead quantum mechanics and gravity would emerge from it.
\item While we would expect the theory to be geometrical in the widest sense, it would not be formulated in any kind of standard spacetime
language, in any dimension.  Probably non-commutative geometry would be important.
\item Given that a fermionic fluid lies at the heart of spacetime, we expect that one theory passes to another via a phase transition, although at the
end of this work (inspired by the work of Murat Gunaydin, Kilian Koepsell and Hermann Nicolai \cite{Gu1}-\cite{Gu3}), we will lay out a possible
series of transitions.
\item Presumably the last transition involves the step from an $\textrm{Spin}(4,4)$-model to a $\U(2,2)$ model and it is exactly at that point that
$\sqrt{-1}$ emerges, together with the concepts and theories of quantum mechanics and space-time. 
\end{itemize}    
We term such a theory a \emph{primordial} theory.  By the last remark, it behooves us to take the group $\textrm{Spin}(4,4)$ seriously and that is what we do in
the present work.  At the time of our becoming aware of the work of Zhang and Hu, we were already heavily involved in the investigation of structures
related to octaves, inspired particularly by the work of John Baez, Tevian Dray and Corinne Manogue \cite{Ba1, Ma2}.  With the relation between
$\textrm{Spin}(4,4)$ and
$\U(2,2)$ also in mind, we had adopted an approach that was neutral with respect to signature, so it was easy to settle on the split signature case. 
Of critical importance to us is the idea of \emph{triality}\cite{Ca1}: we would expect the $\textrm{Spin}(4,4)$ phase to be triality invariant, with no clear
distinction between geometry (points ) and particles (spinors).  Thus we develop the theory in a completely triality invariant way.  In the spirit of Baez (who
explained the octavic theory very beautifully), we focus on the real twenty-seven dimensional Jordan algebra naturally associated to the split
octaves \cite{Jo1}. 
\eject\noindent We analyze from the points of view of geometry: the split analogue of the Moufang octavic plane; and of particle theory: Hamiltonian theory,
Lagrange sub-manifolds and group theory.  Our strategy is close to that of non-commutative geometry: roughly we think of a whole Jordan
algebra element as representing a point.  Some technical remarks: we take as the relevant structures for the Jordan algebra the cross-product and the
determinant.  We do not have use for the trace as such.  In particular we regard the cross-product
$J\times J'$ of elements $J$ and $J'$ of the Jordan algebra, as taking values in the dual Jordan algebra.  Thus although, for completeness,
we do give a formula for
$J\times (J\times J)$, we do not use it later in the work. We are careful to maintain a distinction between the space and its dual throughout.  The
necessity for doing this becomes clear in section ten below, where we see that the cross-product is chiral in nature and switches
chiralities. 
\begin{itemize}\item In sections one and two below, we recall the theory of spinors for $\OO(n,n)$ for the three cases relevant to us, $n =3$, $4$
and
$5$.  The emphasis here is on  explicit computations to back up the more abstract formulas given later.   
\item In section three, we write down our triality axioms.  One should note that these are not confined in principle to any particular dimension:
indeed they might be very interesting in infinite dimension.
\item In section four, we introduce the Jordan algebra, define the cross-product and determinant and develop their basic properties.  
\item In section five we study and solve in general the equation $J\times J = 0$ and contrast our approach with that of Baez.
\item In section six, we discuss the geometrical ideas behind the Jordan algebra.  Briefly the idea is that a "point" should be a non-zero solution of
the equation $J\times J = 0$.  It emerges that a nice way to parametrize the solutions is by the formula $2J = K\times K$, where $\det(K) = 0$.  It is
interesting to note that the projective space of solutions of the equation $J\times J = 0$ is sixteen dimensional (the Moufang projective plane, in
the octavic case) and that the space
$\det(K) = 0$ is twenty-six dimensional, which, using the map $K\rightarrow J = \frac{1}{2}K\times K$, we prove projects surjectively to the
projective  solution space, generically with a ten dimensional fibre: we show that this fibre may be identified with the complement of the null cone in
a flat space of ten dimensions, with a metric of signature $(5,5)$ (in the octavic case the corresponding signature would be $(1,9)$).  \\\\The
numbers
$10$,
$16$ and
$26$ are three of the most important numbers in modern string theory, surely not an accident.  However unlike the Moufang plane, we do not have just the
geometry of points and lines, since it is possible to have linearly independent solutions $(J, J')$ of the equation
$J\times J = J'\times J' = J \times J' = 0$.  We call such solutions "fat points" and their duals "fat lines".  Although for reasons of space and time
we do not discuss the details here, these generalized points have a "string theoretic" interpretation as the image of a suitable curve in the space
$\det(K) = 0$, under the map
$K\rightarrow K\times K$.  
\item In section seven we analyze completely the
solvability of the Jordan algebra equation $K\times K = 2J$, given $J$.  It emerges that the only non-trivial case is the case that $J\times J
= 0$. 
\item In
section eight, we give the full details of the solution of the Jordan algebra equation $K\times K = 2J$, given $J$, such that $J\times J = 0$, keeping
track of the degrees of freedom in the solution.
\item In section nine, we extend our horizons beyond the Jordan algebra to the phase space of Hans Freudenthal, which can be regarded as an augmented
cotangent bundle over the Jordan algebra, augmented with two extra dimensions introduced by Freudenthal, making 56 dimensions in all \cite{Fr1}. We analyze the
Hamiltonian geometry and give quadratic Hamiltonians that generate the group $\E_7$, the symmetry group of the Freudenthal space. We find a beautiful
Lagrangian submanifold on which all our Hamiltonians vanish (including the quartic Hamiltonian invariant of Freudenthal). Following Gunaydin, Koepsell
and Nicolai \cite{Gu1}-\cite{Gu3}, we believe that these results can be extended to $\E_8$, by adding two more dimensions, but we have not yet analyzed this
problem.  
\item In section ten, we develop the Jordan algebra from the point of view of spinors for $\OO(5,5)$.  To do this requires explicitly breaking the
triality structure and this breakdown seems to be a natural avenue for the primordial theory to lose its triality invariance.  
\item In section eleven, we show how to write the triality operations directly in terms of twistors.  Here again the formulas by themselves are not
invariant, even though the underlying structure is.
\item In section twelve we rewrite the formulas of section eleven, more succinctly, using the structure that devolves from
section ten by breaking from $\OO(5,5)$ down to $\OO(2,4)\times \OO(3,1)$, where the $\OO(3,1)$ acts as an "internal symmetry group".  In both
sections eleven and twelve, we can see manifestations of the twistor complex structure: in section eleven, in the
multiplication by the complex number
$x$.  In section twelve, in the multiplication by the Lorentz vector 
$x^a$.  We also discuss the representation of the conformal operators.
\item In section thirteen we present the proposed pattern of symmetry breaking. 
\end{itemize}   
Although we have been concentrating on the relation of the breaking of symmetry to present quantum
theory and spacetime structure, there is one additional payoff that might emerge.  This concerns the development of the concept of
mass.  Building on early work in twistor theory by the first author, Lane Hughston, and Zoltan Perjes \cite{Hu1,Sp6}, the
first author conducted a deep study of the nature of the breaking of conformal invariance.  This led first to a direct construction, in the style of
Eugene Wigner \cite{Wi1}, for the discrete series (and its boundary) for the group
$\SU(2,2)$,  and second to the realization that these representations were bound together in
a single representation of
$\OO(2,6)$ acting as a kind of internal symmetry group \cite{Sp5}.  Following the development outlined here it is natural to ask how this work fits in.   The 
conjecture is that the symmetry breaking applied to the representation of one of our groups will lead to the structure of massive particles found by
the first author.  A candidate is the unitary representation of $\E_8$ found by Gunaydin, Nicolai and Koepsell \cite{Gu3}.    However it is not
yet clear if the decomposition of their representation, with respect to $\SU(2,2)$, involves only positive energy representations; probably not,
since at the
$\SU(2,1)$ level, as discussed by  Gunaydin, Nicolai and Koepsell, representations not belonging to the discrete series of $\SU(2,1)$ occur.  So it
is more likely that some $\E_7$ subrepresentation of their representation is the relevant one, probably one with a minimal value for the Freudenthal
quartic invariant.  However if we proceed on the assumption that at some level only the relevant
$\SU(2,2)$ representations do occur, we are led to ask what the \emph{formula} for the energy momentum operator of the massive particles is.  If we
work at the level of a single Jordan algebra, rather than at the level of the Freudenthal phase space, there seems to be only one reasonable candidate
(in the language of twistor theory, this is the two twistor level: the corresponding $\SU(2,2)$ representations all lie at the boundary of the discrete
series; for the generic discrete series representation, at least three twistors are needed).  The relevant formula is given and discussed at the end of
section twelve.  We emphasize however that despite its attractive appearance, we do not presently have a clear-cut rationale for the formula.

\eject\noindent
\section{Spinors for  $\mathcal{O}(n,n)$} 
We work with vector spaces over the reals, of finite dimension, unless otherwise specified.  If $\V$ is any vector space, we denote by $I_\V$ the
identity endomorphism of
$\V$. We denote the dual space of $\V$ by $\V^*$ and we identify $\V$
and $(\V^*)^*$.  Suppose that
$\V$ is a real vector space, of positive dimension
$2n$, equipped with a symmetric bilinear form $g$ of type $(n,n)$. The symmetry group of $\V$ is isomorphic to the pseudo-orthogonal group
$\mathcal{O}(n,n)$.  The spin representation $\mathcal{S}$ of the Clifford algebra of $(\V, g)$ has dimension $2^n$ over the
reals.  The space $\V$ acts on $\mathcal{S}$, such that $z^2$ acts as
$g(z,z)$ times the identity operator, for any $z \in \V$. If
$\{e_i: 1\le i\le 2n\}$ is an orthonormal frame for $\V$, such that $g(e_i, e_i) = 1$
for
$1\le i\le n$ and $g(e_i, e_i) = -1$ for $n<i\le 2n$, then there exists a basis for
$\mathcal{S}$, such that $e_i$ is symmetric for $1\le i \le n$ and skew for $n<i\le 2n$. 
Put $\alpha = e_1e_2\dots e_n$,
$\beta = e_{n+1}e_{n+2}\dots e_{2n}$ and $\gamma = \alpha\beta$.  Then we have the
relations:
\[\alpha^T = (-1)^{\frac{1}{2}n(n-1)}\alpha,\hspace{10pt} \beta^T =
(-1)^{\frac{1}{2}n(n+1)}\beta,\hspace{10pt} \gamma^T = \gamma,\] 
\[\alpha^2 = 
(-1)^{\frac{1}{2}n(n-1)}I_{\mathcal{S}}, \hspace{10pt}\beta^2 = (-1)^{\frac{1}{2}n(n+1)}I_{\mathcal{S}},
\hspace{10pt}\gamma^2 = I_{\mathcal{S}},\]
\[ \beta\hspace{-1pt}\gamma\hspace{-2.5pt} =\hspace{-2.5pt} (-1)^{n}\hspace{-1pt}\gamma\hspace{-1pt}\beta \hspace{-2.5pt}=\hspace{-2.5pt}
(-1)^{\frac{1}{2}n(n\hspace{-1pt}-\hspace{-1pt}1)}\hspace{-1pt}\alpha, \hspace{4pt}\alpha\hspace{-1pt}\beta \hspace{-2.5pt}=\hspace{-2.5pt}
(-1)^n\hspace{-1pt}\beta\alpha\hspace{-2.5pt} =\hspace{-2.5pt} \gamma,
\hspace{4pt}\gamma\hspace{-1pt}\alpha\hspace{-2.5pt} =\hspace{-2.5pt} (-1)^{n}\alpha\hspace{-1pt}\gamma\hspace{-2.5pt} =
\hspace{-2.5pt}(-1)^{\frac{1}{2}n(n\hspace{-1pt}+\hspace{-1pt}1)}\hspace{-1pt}\beta. \] For each spinor $\psi$, put
$\psi'_\alpha =
(\alpha\psi)^T$ and $\psi'_\beta = (\beta\psi)^T$.  Then the mappings $\psi\rightarrow \psi'_\alpha$
and $\psi\rightarrow \psi'_\beta$ give canonical maps from the spin space $\mathcal{S}$ to its
dual $\mathcal{S}^*$. These maps are related by the formula
$\psi'_\beta = (-1)^{\frac{1}{2}n(n+1)}\psi'_\alpha\gamma$.  The dual pairings $(\psi,
\phi)_\alpha =
\psi'_\alpha\phi$ and $(\psi, \phi)_\beta =
\psi'_\beta\phi$, defined for any spinors
$\phi$ and
$\psi$ are:
\begin{itemize}\item Both non-degenerate symmetric in the case $n = 0$ mod $4$.
\item Non-degenerate symmetric for $(\psi,
\phi)_\alpha $ and  symplectic for $(\psi, \phi)_\beta $ in the case $n = 1$ mod $4$.
\item Both
symplectic in the case $n = 2$ mod $4$.
\item Non-degenerate symmetric for $(\psi,
\phi)_\beta $ and  symplectic for $(\psi, \phi)_\alpha $ in the case $n = 3$ mod $4$.
\end{itemize}
Multiplication by $\gamma$ is a canonical linear operator on the
spin space.  
The half-spin spaces $\mathcal{S}^{\pm}$ are the two eigenspaces of $\gamma$, with
$\pm \gamma$ the identity operator on $\mathcal{S}^{\pm}$.  Each of $\mathcal{S}^{+}$ and
$\mathcal{S}^-$ has dimension
$2^{n-1}$ over the reals.  Multiplication by $z\in \V$ maps $\mathcal{S}^\pm$ to
$\mathcal{S}^\mp$.
\begin{itemize}\item For $n$ even, the spaces $\mathcal{S}^{+}$ and
$\mathcal{S}^-$ are mutually orthogonal with respect to both the pairings  $(\psi,
\phi)_\alpha$ and $(\psi, \phi)_\beta$ and the restrictions of these pairings to  
$\mathcal{S}^{+}$ and $\mathcal{S}^-$ coincide up to a sign, giving each a metric structure
when
$n = 0$ mod $4$ and a symplectic structure when $n = 2$ mod $4$.
\item For $n$ odd, the restrictions of the pairings  $(\psi,
\phi)_\alpha$ and $(\psi, \phi)_\beta$ to the subspaces $\mathcal{S}^{+}$ and
$\mathcal{S}^-$ vanish identically.  Also each of these pairings sets up a natural duality
between
$\mathcal{S}^{+}$ and
$\mathcal{S}^-$ and the induced dualities coincide up to a sign.

\end{itemize}
Associated to the spin representation of $\V$ are two real trilinear forms, $(\psi, z, \phi)_\alpha$ and $(\psi, z,
\phi)_\beta$, defined as follows, for any spinors
$\psi$ and $\phi$ and for any vector $z$:  
\[  (\psi, z, \phi)_\alpha
= (\psi, z\phi)_\alpha = (-1)^{\frac{1}{2}(n+2)(n - 1)}(\phi, z, \psi)_\alpha, \]\[
(\psi, z,
\phi)_\beta = (\psi, z\phi)_\beta = (-1)^{\frac{1}{2}n(n-1)}(\phi, z, \psi)_\beta.\]  These two trilinear forms carry the same
information.
\begin{itemize} \item In the cases $n = 0$ and $n = 2$ mod $4$, each trilinear form vanishes if
$\psi$ and $\phi$ belong to the same half-spin space, so may be considered as a form on
$\mathcal{S}^+\times \V\times \mathcal{S}^-$.  Then the two trilinear forms agree up to a
sign, so it suffices to focus on one only.  Usually we use the form that is symmetric in its spinor arguments: so the form $(\psi, z\phi)_\beta$ in the
case
$n = 0$ mod
$4$ and the form $(\psi, z\phi)_\alpha$ in the case $n = 2$ mod $4$. 
\item In the cases $n = 1$ and $n = 3$ mod $4$, each trilinear form vanishes if
$\psi$ and $\phi$ belong to distinct half-spin spaces, so the information in the trilinear
form reduces to that in the restrictions of the form to the spaces $\mathcal{S}^+\times
\V\times \mathcal{S}^+$ and $\mathcal{S}^-\times \V\times \mathcal{S}^-$.  Again it
suffices to focus on one trilinear form only.  The trilinear forms are both  symmetric in their spinor arguments if $n = 1 $ mod $4$ and skew in their
spinor arguments if $n = 3 $ mod $4$.
\end{itemize}
The maps $\phi\rightarrow \phi'_\alpha$ and $\phi \rightarrow \phi'_\beta$ extend naturally to anti-automorphisms of the Clifford algebra, such that
$(z\phi)'_\alpha\psi = \phi'_\alpha z'_\alpha\psi$   and $(z\phi)'_\beta\psi = \phi'_\beta z'_\beta\psi$, for any vector $z$ and any spinors $\psi$
and $\phi $.  Then we have $z'_\alpha = (-1)^{n-1} z$ and $z'_\beta = (-1)^n z$, for any vector $z$.   
\eject\noindent
\section{Spinors for  $\mathcal{O}(n,n)$: the cases $n = 3$, $4$ and $5$} 
We first recall the algebra of quaternions, $\HH$.  For any quaternion $q$, $\overline{q}$ denotes its quaternion conjugate.  Then $q$ is
real iff
$q =
\overline{q}$, whereas $q$ is pure imaginary iff $q = - \overline{q}$. We repeatedly use the
facts that
$q\overline{q} =
\overline{q}q$ is real (positive if
$q\ne 0$: we put $|q|=\sqrt{q\overline{q}}$), that $q$ commutes under multiplication by any
real and that the following identities hold, for all quaternion pairs, $(q,r)$:
\[ \overline{\overline{q}} =  q, \hspace{10pt}\overline{(qr)} = \overline{r}\hspace{2pt}\overline{q}, \hspace{10pt}qr - rq =
\overline{q}\hspace{2pt}\overline{r}  -
\overline{r}\hspace{2pt}\overline{q}.\]
To analyze the $\mathcal{O}(n,n)$ spinors for the cases $n = 3$, $4$ and
$5$, we will use quaternions to build the spinor spaces $\mathcal{S}$ and the vector space $\V$.
\begin{itemize}\item For $\mathcal{O}(3,3)$, the spin space $\mathcal{S}$ is $\HH^2$, all quaternion pairs.  The vector space $\V$ is then the
collection of all pairs $(x, y)\in\HH^2$ with $x$ and $y$ imaginary, with the metric $|(x, y)|^2 = y^2 - x^2$.  Each half-spin space $\mathcal{S}^\pm$
is an $\HH$.
\item
For $\mathcal{O}(4,4)$, the spin space $\mathcal{S}$ is $\HH^4$, all quaternion quartets. The vector space $\V$ comprises all quartets
$(u, v, x, y)\in \R^2\times \HH^2$ with $x$ and $y$ imaginary, with the metric $|(u ,v, x,
y)|^2 = uv + x^2 - y^2$.   Each $\mathcal{S}^\pm$ is an $\HH^2$.
\item 
For $\mathcal{O}(5,5)$, the spin space $\mathcal{S}$ is $\HH^8$, all quaternion octets.  The vector space $\V$ comprises all sextuples
$(s,\hspace{-1.5pt} t, \hspace{-1.5pt}u,\hspace{-1.5pt} v,\hspace{-1.5pt} x,\hspace{-1.5pt} y)\in \R^4\times \HH^2$ with $x$ and $y$ imaginary, with
the metric
$|(s,\hspace{-1.5pt} t, \hspace{-1.5pt}u,\hspace{-1.5pt} v,\hspace{-1.5pt} x,\hspace{-1.5pt} y)|^2\hspace{-1.5pt} =\hspace{-1.5pt} st
\hspace{-1.5pt}-\hspace{-1.5pt}uv\hspace{-1.5pt} -\hspace{-1.5pt} x^2
\hspace{-1.5pt}+
\hspace{-1.5pt}y^2$. Each
$\mathcal{S}^\pm$ is an $\HH^4$.
\end{itemize} 
We now consider  the details of each case.
\begin{itemize}\item Spinors for
$\mathcal{O}(3,3)$ may be given as the space $\HH^2$, all pairs of quaternions: $(a,b)$. For a given quaternion pair
$(x,y)$, consider the endomorphisms $\gamma^\pm_{(x,y)}$ of $\HH$ and  $\gamma_{(x,y)}$ of $\HH^2$, defined as follows:
\[ \gamma^+_{(x,y)}(a) =
\overline{a}y + x\overline{a},\hspace{10pt} \gamma^-_{(x,y)}(a) = \overline{a}x -
y\overline{a},
\hspace{10pt}\gamma^-_{(x,y)}= \gamma^+_{-y, x}.\]
\[ \gamma_{(x,y)}(a,b) = (\overline{b}x - y\overline{b}, \overline{a}y +
x\overline{a}) = (\gamma^-_{(x,y)}(b), \gamma^+_{(x,y)}a), \]
Here $a$ and $b$ are arbitrary quaternions.  Then we have the relation:
\[ \gamma^2_{(x,y)}(a,b) = (x\overline{x} - y\overline{y})(a, b) +
(\overline{y}ax - ya\overline{x}, \overline{x}by - xb\overline{y}).\]
In particular, if $x$ and $y$ are both \emph{imaginary} quaternions, as we shall assume
henceforth, we have the desired spin representation for $\mathcal{O}(3,3)$: 
\[ \gamma^+_{(x,y)}\gamma^-_{(x,y)} = \gamma^-_{(x,y)}\gamma^+_{(x,y)} = (y^2 - x^2)I_{\HH}.\]
\[ \gamma^2_{(x,y)} = (x\overline{x} - y\overline{y})I_{\HH^2} = (y^2 - x^2)I_{\HH^2},\]
\eject\noindent
The transpose of $\psi = (a,b)$ is $\psi^T = (\overline{a},\overline{b})$, with $2\psi^T\phi
= \overline{a}c + \overline{c}a + \overline{b}d + \overline{d}b$, where $\phi = (c, d)$.  The
invariant structure operators may be taken as follows:
$\alpha = \gamma_{(i, 0)}\gamma_{(j, 0)}\gamma_{(k, 0)}$ and 
$\beta =
\gamma_{(0, i)}\gamma_{(0, j)}\gamma_{(0, k)}$ and $\gamma = \alpha\beta$.  We have:
\[ \alpha(a,b) =  (-\overline{b}, \overline{a}), \hspace{10pt}\beta(a, b) = (\overline{b},
\overline{a}),\hspace{10pt} \gamma(a,b) = (-a,b).\]
The canonical dual of
$\psi = (a,b)$ is $\psi' = (\beta\psi)^T = (b, a)$.  The canonical dual pairing is
$2\psi'\phi = ad + \overline{da} + bc + \overline{c}\overline{b}$. The half-spin spaces are the
eigenstates of
$\gamma$ so are the spinors of the form
$(a,0)$ or
$(0, b)$.  The associated real trilinear form is:
\[  2\phi'\gamma_{x,y}\psi =
2(d,c).(\overline{b}x - y\overline{b}, \overline{a}y +
x\overline{a}) \]\[= d\overline{b}x - dy\overline{b} -
xb\overline{d} + by\overline{d} + c\overline{a}y + cx\overline{a}- ya\overline{c} -
ax\overline{c} \]
\[= d\overline{b}x - \overline{b}dy -
xb\overline{d} + y\overline{d}b + c\overline{a}y + \overline{a}cx- ya\overline{c} -
x\overline{c}a \]
\[ = (x, y).(b\overline{d} - d\overline{b} + \overline{c}a -
\overline{a}c, \overline{b}d -\overline{d}b + a\overline{c} - c\overline{a}).\] 
The norm squared of the vector $(b\overline{d} - d\overline{b} + \overline{c}a -
\overline{a}c, \overline{b}d -\overline{d}b + a\overline{c} - c\overline{a})$
is:
\[-(b\overline{d} - d\overline{b} + \overline{c}a -
\overline{a}c)^2 + (\overline{b}d -\overline{d}b + a\overline{c} - c\overline{a})^2\]
\[ = - (b\overline{d}
-d\overline{b})(\overline{c}a -\overline{a}c) - (\overline{c}a -\overline{a}c)(b\overline{d}
-d\overline{b}) \]\[ - (\overline{d}b -\overline{b}d)(a\overline{c} - c\overline{a})
- (a\overline{c} - c\overline{a})(\overline{d}b - \overline{b}d)   \]
\[ = - b\overline{d}\overline{c}a + b\overline{d}\overline{a}c +d\overline{b}\overline{c}a
- d\overline{b}\overline{a}c - \overline{c}ab\overline{d} + \overline{a}cb\overline{d} +
\overline{c}ad\overline{b} -
\overline{a}cd\overline{b}\]\[  - \overline{d}ba\overline{c}
+\overline{b}da\overline{c} + \overline{d}bc\overline{a}
-\overline{b}dc\overline{a} - a\overline{c}\overline{d}b + c\overline{a}\overline{d}b +
a\overline{c}\overline{b}d - c\overline{a}\overline{b}d\]
\[ = 2(- a\overline{c}\overline{d}b - ab\overline{d}\overline{c} + ad\overline{b}\overline{c} +
a\overline{c}\overline{b}d +
 \overline{d}bc\overline{a} + cb\overline{d}\overline{a} -
cd\overline{b}\overline{a} -\overline{b}dc\overline{a})\]
If any one of the four component half-spinors vanishes, this norm vanishes and the vector is
null.
\item Using the operators $\gamma^{\pm}_{x,y}$ we now construct the spin representation for
$\mathcal{O}(4,4)$:  the new operator $\gamma_{u,v,x,y}$ is given by the matrix:
\[ \gamma_{u,v,x,y} =\hspace{3pt}
\begin{array}{|cccc|}0&0&u&\gamma^-_{x,y}\\0&0&\gamma^+_{x,y}&v\\v&-\gamma^-_{x,y}&0&0\\-\gamma^+_{x,y}&u&0&0\end{array}\hspace{3pt} =
\hspace{3pt}\begin{array}{|cc|}0&\gamma^+_{u,v,x,,y}\\\gamma^-_{u,v,x,y}&0\end{array}\hspace{3pt}.\] Here $u$ and $v$ are real numbers and $x$ and $y$
are imaginary quaternions.  Then $\gamma_{u,v,x,y}$ acts on a spinor of the form $(a,b,c,d)\in
\HH^4$, written as a column vector.  Explicitly we have:
\[   \gamma^+_{u,v,x,y}\hspace{3pt}\begin{array}{|c|}c\\d\end{array}\hspace{7pt}
=\hspace{7pt}\begin{array}{|c|} uc +\overline{d}x - y\overline{d}\\vd + \overline{c}y +
x\overline{c}\end{array}\hspace{3pt}, \hspace{20pt} \gamma^-_{u,v,x,y}\hspace{3pt}\begin{array}{|c|}a\\b\end{array}\hspace{7pt}
=\hspace{7pt}\begin{array}{|c|}va - \overline{b}x + y\overline{b}\\ub - \overline{a}y -
x\overline{a}\end{array}\hspace{3pt}, \]
\[ \gamma_{u,v,x,y}\hspace{3pt}\begin{array}{|c|}a\\b\\c\\d\end{array}\hspace{7pt}
=\hspace{7pt}\begin{array}{|c|} uc +\overline{d}x - y\overline{d}\\vd + \overline{c}y +
x\overline{c}\\va - \overline{b}x + y\overline{b}\\ub - \overline{a}y -
x\overline{a}\end{array}\hspace{3pt}.\]
It is easily checked that the operators $\gamma^\pm_{u,v,x,y}$ and 
$\gamma_{u,v,x,y}$ obey the required  relations:
\[  \gamma^+_{u,v,x,y}\gamma^-_{u,v,x,y} = \gamma^-_{u,v,x,y}\gamma^+_{u,v,x,y} = (uv + x^2 -
y^2)I_{\HH^2},
\hspace{10pt}\gamma_{u,v,x,y}^2 = (uv + x^2 - y^2)I_{\HH^4}.\]
The operators $\alpha$, $\beta$ and $\gamma$ can be constructed as follows:
\[ \alpha = \gamma_{1,1,0,0}\gamma_{0,0,0,i}\gamma_{0,0,0,j}\gamma_{0,0,0,k}, \hspace{10pt}
\beta =
\gamma_{-1,1,0,0}\gamma_{0,0,i,0}\gamma_{0,0,j,0}\gamma_{0,0,k,0}, \hspace{10pt}\gamma = \alpha\beta\]
\[\alpha\hspace{3pt}\begin{array}{|c|}a\\b\\c\\d\end{array}\hspace{7pt}
=
\hspace{7pt}\begin{array}{|c|}\overline{b}\\\overline{a}\\-\overline{d}\\-\overline{c}\end{array}\hspace{7pt},
\hspace{10pt}\beta\hspace{3pt}\begin{array}{|c|}a\\b\\c\\d\end{array}\hspace{7pt}
=
\hspace{7pt}\begin{array}{|c|}\overline{b}\\\overline{a}\\\overline{d}\\\overline{c}\end{array}\hspace{7pt},\hspace{10pt}\gamma\hspace{3pt}\begin{array}{|c|}a\\b\\c\\d\end{array}\hspace{7pt}
=
\hspace{7pt}\begin{array}{|c|}a\\b\\-c\\-d\end{array}\hspace{7pt}.\]
The transpose of $\psi = (a,b,c,d)$ is $\psi^T = (\overline{a}, \overline{b}, \overline{c},
\overline{d})$.  The dual pairing of $\psi$ with $\phi = (e,f,g,h)\in\HH^4$ is given by:
\[ 2\psi^T\phi = \overline{a}e + \overline{b}f + \overline{c}g + \overline{d}h + \overline{e}a
+ \overline{f}b + \overline{g}c + \overline{h}d. \]
We put $\psi' = (\beta\psi)^T = (b,a,d,c)$, giving the canonical
dual pairing:
\[ 2\psi'\phi = be + af + dg + ch +  \overline{h}\overline{c} + \overline{g}\overline{d} +
\overline{f}\overline{a} + \overline{e}\overline{b}.\] 
Then the symmetric trilinear form is given by the following:
\[ \psi'\gamma_{u,v,x,y}\psi = \frac{1}{2}(u(bc + cb + \overline{b}\overline{c} + \overline{c}\overline{b}) + v(ad + da +
\overline{a}\overline{d} + \overline{d}\overline{a}) + x(\overline{c}a -\overline{a}c + b\overline{d} - d\overline{b})\]\[ +
y(a\overline{c} - c\overline{a} + \overline{b}d - \overline{d}b) + (a\overline{c} - c\overline{a} + \overline{b}d - \overline{d}b)y +
(\overline{c}a -\overline{a}c + b\overline{d} - d\overline{b})x)\]
\[ = (u,v,x,y).(\psi'\underline{\gamma}\psi), \]
\[ \psi'\underline{\gamma}\psi = (2(ad +  \overline{d}\overline{a}), \hspace{3pt} 2(bc +  \overline{c}\overline{b}),
\hspace{3pt}
\overline{c}a -\overline{a}c + b\overline{d} - d\overline{b}, \hspace{3pt}c\overline{a} - a\overline{c} + \overline{d}b -
\overline{b}d).\] 
The norm squared of this latter vector is calculated as follows:
\[ | (\psi'\underline{\gamma}\psi)|^2 = 4(ad + \overline{d}\overline{a})(bc + \overline{c}\overline{b}) + (\overline{c}a -\overline{a}c +
b\overline{d} - d\overline{b})^2  - (c\overline{a} - a\overline{c} + \overline{d}b -
\overline{b}d)^2\] 
\[ = 4(ad + \overline{d}\overline{a})(bc + \overline{c}\overline{b}) + (\overline{c}a -\overline{a}c)(b\overline{d} - d\overline{b}) +
(b\overline{d} - d\overline{b})(\overline{c}a -\overline{a}c)\]\[  - (c\overline{a} - a\overline{c})(\overline{d}b -
\overline{b}d) - (\overline{d}b - \overline{b}d) (c\overline{a} - a\overline{c}) = 2aU + 2\overline{U}\overline{a}, \]
\[ U = 2dbc + 2d\overline{c}\overline{b} - d\overline{b}\overline{c} + b\overline{d}\overline{c} + \overline{c}\overline{d}b -
\overline{c}\overline{b}d\]
\[  = 2dcb + d\overline{b}\overline{c}  + b\overline{d}\overline{c} + \overline{c}\overline{d}b -
\overline{c}\overline{b}d\]
\[  = 2(dc + \overline{c}\overline{d})b + (d\overline{b}  + b\overline{d})\overline{c} - \overline{c}(\overline{d}b
+\overline{b}d) = 2b(cd + \overline{d}\overline{c}), \]
\[ | (\psi'\underline{\gamma}\psi)|^2 = 4(ab + \overline{b}\overline{a})(cd + \overline{d}\overline{c}).\]
In particular if any of the quantities $a$, $b$, $c$ or $d$ vanishes, then the vector $\psi'\underline{\gamma}\psi$ is null.\\\\
We may now reformulate: introduce spinors $\alpha \in \mathcal{S}^+$ and $\gamma\in \mathcal{S}^-$ and the vector $\beta\in\V$ as follows:
\[ \alpha = \hspace{3pt}\begin{array}{|c|}a\\b\end{array} \hspace{3pt}\in \mathcal{S}^+, \hspace{10pt} \gamma =
\hspace{3pt}\begin{array}{|c|}c\\d\end{array} \hspace{3pt}\in \mathcal{S}^-, \hspace{10pt}\beta = (u,v,x,y) \in \V.\]
Here $u$ and $v$ are real numbers, whereas $a$, $b$, $c$, $d$, $x$ and $y$ are quaternions, with $x$ and $y$ pure imaginary.
Each of space is given a natural $(4,4)$ metric:
\[ \alpha^2 = \alpha.\alpha = 2ab + 2\overline{b}\overline{a}, \hspace{10pt} \gamma^2 = \gamma.\gamma = 2cd + 2\overline{d}\overline{c},
\hspace{10pt}\beta^2 = \beta.\beta = uv + x^2 - y^2.\]
Then we introduce three products and a triple product:
\[ \V \times \mathcal{S}^-\rightarrow \mathcal{S}^+,\] \[ (\beta, \gamma) \rightarrow \beta\gamma =\hspace{3pt}
\gamma^+_{u,v,x,y}\hspace{3pt}\begin{array}{|c|}c\\d\end{array}\hspace{3pt} = 
\hspace{3pt}\begin{array}{|c|}uc + \overline{d}x - y\overline{d}\\vd + \overline{c}y + x\overline{c}\end{array}\hspace{3pt}, \]
\[  \mathcal{S}^- \times \mathcal{S}^+\rightarrow \V,\] \[ (\gamma, \alpha) \rightarrow \gamma\alpha = ( 2(ad + 
\overline{d}\overline{a}), \hspace{3pt} 2(bc +  \overline{c}\overline{b}),
\hspace{3pt}
\overline{c}a -\overline{a}c + b\overline{d} - d\overline{b}, \hspace{3pt}c\overline{a} - a\overline{c} + \overline{d}b -
\overline{b}d),\]
\[  \mathcal{S}^+\times \V \rightarrow \mathcal{S}^-, \]\[ (\alpha, \beta) \rightarrow \alpha\beta =\hspace{3pt}
\gamma^-_{u,v,x,y}\hspace{3pt}\begin{array}{|c|}a\\b\end{array}\hspace{3pt} = 
\hspace{3pt}\begin{array}{|c|}va - \overline{b}x + y\overline{b}\\ub - \overline{a}y - x\overline{a}\end{array}\hspace{3pt},\]
\[  \mathcal{S}^+\times \V \times \mathcal{S}^-\rightarrow \R,\]\[ (\alpha, \beta, \gamma)\rightarrow (\alpha\beta\gamma) = \alpha.(\beta\gamma) = \]
\[ \frac{1}{2}(u(bc + cb + \overline{b}\overline{c} + \overline{c}\overline{b})
+ v(ad + da +
\overline{a}\overline{d} + \overline{d}\overline{a}) + x(\overline{c}a -\overline{a}c + b\overline{d} - d\overline{b})\]\[ +
y(a\overline{c} - c\overline{a} + \overline{b}d - \overline{d}b) + (a\overline{c} - c\overline{a} + \overline{b}d - \overline{d}b)y +
(\overline{c}a -\overline{a}c + b\overline{d} - d\overline{b})x).\]
We also write $\alpha\beta = \beta\alpha$, $\beta\gamma = \gamma\beta$ and $\gamma\alpha = \alpha\gamma$.  Then by direct calculation, we
may verify the following relations:
\[ (\beta\gamma)^2 = \beta^2 \gamma^2, \hspace{10pt} (\gamma\alpha)^2 = \gamma^2\alpha^2, \hspace{10pt} (\alpha\beta)^2 =
\alpha^2\beta^2,\]
\[ \alpha(\alpha\beta) = \alpha^2\beta, \hspace{10pt} \alpha(\alpha\gamma) = \alpha^2\gamma,\]
\[ \beta(\beta\gamma) = \beta^2\gamma, \hspace{10pt} \beta(\beta\alpha) = \beta^2\alpha,\]
\[ \gamma(\gamma\alpha) = \gamma^2\alpha, \hspace{10pt} \gamma(\gamma\beta) = \gamma^2\beta,\]
\[ (\alpha\beta\gamma) = \alpha.(\beta\gamma) = \beta.(\gamma\alpha) = \gamma.(\alpha\beta).\]
By the last formula, it is legitimate to regard the trilinear form as being symmetric under permutation of its arguments.   Also by the time the last
formula arrives, we see that we have constructed a completely symmetric structure.  However the formulas themselves are
not manifestly symmetric: indeed $\V$ is on an apparently different footing from $\mathcal{S}^\pm$.  Note that we may select three elements of
these spaces, $\epsilon_0\in\V$, $\epsilon_\pm\in \mathcal{S}^\pm$, each of unit length, that together behave as an identity operator:
\[\epsilon_0 = (1,1,0,0), \hspace{10pt}\epsilon_{+} =  \hspace{3pt}\begin{array}{|c|}\frac{1}{2}\\\frac{1}{2}\end{array} \hspace{3pt}, 
\hspace{10pt}\epsilon_{-} = \hspace{3pt}
\begin{array}{|c|}\frac{1}{2}\\\frac{1}{2}\end{array} \hspace{3pt},\]
\[ \epsilon_0^2 = \epsilon_{\pm}^2 = 1, \hspace{10pt}\epsilon_0\epsilon_{\pm} = \epsilon_{\mp}, \hspace{10pt} \epsilon_+\epsilon_- = \epsilon_0,
\hspace{10pt}(\epsilon_0\epsilon_+\epsilon_-) = 1.\]
Using these identitites, all three multiplication rules may be shown to be equivalent.  We may make each of $\V$ and $\mathcal{S}^\pm$ into algebras by
the multiplication rules:
$\beta\beta' = (\epsilon_+\beta)(\epsilon_{-}\beta')$, $\alpha\alpha' = (\epsilon_0\alpha)(\epsilon_{-}\alpha')$ and $\gamma\gamma' =
(\epsilon_0\gamma)(\epsilon_{+}\gamma')$, defined for any elements $\beta$ and
$\beta'$ of
$\V$, $\alpha$ and $\alpha'$ of $\mathcal{S}^+$ and $\gamma$ and $\gamma'$ of $\mathcal{S}^-$.  Then $\epsilon_0$ and $\epsilon_\pm$ are the identities
for $\V$, $\mathcal{S}^\pm$, respectively.  Also multiplication by, for example $\epsilon_+$ gives an isomorphism of the $\mathcal{S}^-$ algebra with
the
$\V$ algebra  and vice-versa.  Each such algebra is then isomorphic to the algebra of the split octaves.
\item Finally we move to the case of $\mathcal{O}(5,5)$.  We use the operators $\gamma^\pm_{u,v,x,y}$ constructed for the $\mathcal{O}(4,4)$ case to
produce the required operator $\gamma_{s,t,u,v,x,y}$, defined for $(s,t,u,v)\in\R^4$ and for $x$ and $y$ pure imaginary quaternions:
\[ \gamma_{s,t,u,v,x,y} =
\hspace{3pt}\begin{array}{|cccc|}0&0&s&\gamma^+_{u,v,x,y}\\0&0&\gamma^-_{u,v,x,y}&t\\t&-\gamma^+_{u,v,x,y}&0&0\\-\gamma^-_{u,v,x,y}&0&0\end{array}\hspace{3pt}.\]
Explicitly, we have:
\[ \gamma_{s,t,u,v,x,y}\hspace{3pt}\begin{array}{|c|}a\\b\\c\\d\\e\\f\\g\\h\end{array}\hspace{7pt} =
\hspace{7pt}\begin{array}{|c|}se + ug + \overline{h}x - y\overline{h}\\sf + vh + \overline{g}y + x\overline{g}\\tg + ve - \overline{f}x +
y\overline{f}\\th + uf - \overline{e}y - x\overline{e}\\ta - uc - \overline{d}x + y\overline{d}\\tb - vd - \overline{c}y - x\overline{c}\\sc - va +
\overline{b}x - y\overline{b}\\sd - ub + \overline{a}y + x\overline{a} \end{array}.\]
Here the spinor $(a,b,c,d,e,f,g,h)\in\HH^8$. It is straightforward to verify the required relation:
\[ \gamma^2_{s,t,u,v,x,y} = (st - uv - x^2 + y^2)I.\]
We have for the operators $\alpha$, $\beta$ and $\gamma$:
\[ \alpha = \gamma_{1,1,0,0,0,0}\gamma_{0,0,-1,1,0,0}\gamma_{0,0,0,0,i,0}\gamma_{0,0,0,0,j,0}\gamma_{0,0,0,0,k,0}\]
\[ \beta = \gamma_{-1,1,0,0,0,0}\gamma_{0,0,1,1,0,0}\gamma_{0,0,0,0,0,i}\gamma_{0,0,0,0,0,j}\gamma_{0,0,0,0,0,k}\]
\[ \gamma = \alpha\beta.\]
Explicitly, we have:
\[ \alpha\hspace{3pt}\begin{array}{|c|}a\\b\\c\\d\\e\\f\\g\\h\end{array}\hspace{7pt} =
\hspace{7pt}\begin{array}{|c|}\overline{f}\\\overline{e}\\\overline{h}\\\overline{g}\\\overline{b}\\\overline{a}\\\overline{d}\\\overline{c}\end{array}\hspace{3pt},\hspace{16pt}
\beta\hspace{3pt}\begin{array}{|c|}a\\b\\c\\d\\e\\f\\g\\h\end{array}\hspace{7pt} =
\hspace{7pt}\begin{array}{|c|}-\overline{f}\\-\overline{e}\\-\overline{h}\\-\overline{g}\\\overline{b}\\\overline{a}\\\overline{d}\\\overline{c}\end{array}\hspace{3pt},\hspace{16pt}
\gamma\hspace{3pt}\begin{array}{|c|}a\\b\\c\\d\\e\\f\\g\\h\end{array}\hspace{7pt} =
\hspace{7pt}\begin{array}{|c|}a\\b\\c\\d\\-e\\-f\\-g\\-h\end{array}\hspace{7pt}\hspace{3pt}.
\]
The transpose of $\psi = (a,b,c,d,e,f,g,h)$ (written as a column vector) is the row $\psi^T = (\overline{a}, \overline{b}, \overline{c}, \overline{d},
\overline{e}, \overline{f}, \overline{g}, \overline{h})$, with the dual pairing $\psi^T\phi$ for $\phi = (j,k,l,m,n,p,q,r)$, given by the
formula:
\[ 2\psi^T\phi \hspace{0pt}= \hspace{0pt}\overline{a}j+ \overline{b}k\hspace{-2pt} +\hspace{-2pt}
\overline{c}l+
\overline{d}m + 
\overline{e}n + \overline{f}p +\overline{g}q  + \overline{h}r
\]\[+ 
\overline{r}h + \overline{q}g +
\overline{p}f +
\overline{n}e + 
\overline{m}d + \overline{l}c+ \overline{k}b + \overline{j}a.\]
Then the invariant scalar product $\psi'\phi  = (\alpha\psi)^T\phi$ is given as follows:
\[ 2\psi'\phi = fj+ek+hl+gm
+bn +ap +dq +cr \]\[+ 
\overline{r}\hspace{1.5pt}\overline{c} + \overline{q}\overline{d}+
\overline{p}\overline{a}+
\overline{n}\overline{b} + 
\overline{m}\hspace{1.5pt}\overline{g} + \overline{l}\overline{h}+ \overline{k}\overline{e}
+
\overline{j}\overline{f}.\]
In particular, we have: $\psi'\psi = \hspace{-2pt}af\hspace{-2pt} +\hspace{-2pt}be\hspace{-2pt} +\hspace{-2pt}ch\hspace{-2pt} +\hspace{-2pt}dg
\hspace{-2pt}+ \overline{g}\overline{d} \hspace{-2pt}+\hspace{-2pt} \overline{h}\overline{c} \hspace{-2pt}+\hspace{-2pt} \overline{e}\overline{b}
\hspace{-2pt}+\hspace{-2pt}
\overline{f}\overline{a}$.
Then for the triple product $\psi'\gamma_{s,t,u,v,x,y}\psi$, we have the expresssion:
\[ \psi'\gamma_{s,t,u,v,x,y}\psi = 2s(cd + ef + \overline{f}\overline{e} + \overline{d}\overline{c}) + 2t(ab + gh + \overline{h}\overline{g} +
\overline{b}\overline{a})\]\[ + 2u(fg - bc - \overline{c}\overline{b} + \overline{g}\overline{f}) + 2v(eh- ad - \overline{d}\overline{a} +
\overline{h}\overline{e})\]
\[ + x(\overline{a}c - \overline{c}a - b\overline{d} + d\overline{b} -\overline{e}g + \overline{g}e + f\overline{h} - h\overline{f}) \]\[+
y(c\overline{a} - a\overline{c} +\overline{d}b -\overline{b}d -g\overline{e} + e\overline{g} + \overline{f}h - \overline{h}f)\]
\[ + (\overline{a}c - \overline{c}a - b\overline{d} + d\overline{b} -\overline{e}g + \overline{g}e + f\overline{h} - h\overline{f})x\]
\[  + (c\overline{a} - a\overline{c} +\overline{d}b -\overline{b}d -g\overline{e} + e\overline{g} + \overline{f}h - \overline{h}f)y.\]
\[ = (s,t,u,v,x,y).(\psi'\underline{\gamma}\psi), \]
\[ \frac{1}{2}\psi'\underline{\gamma}\psi = (2(ab + gh + \overline{h}\overline{g} +
\overline{b}\overline{a}),\hspace{10pt} 2(cd + ef + \overline{f}\overline{e} + \overline{d}\overline{c}), \]\[ - 2(eh- ad -
\overline{d}\overline{a} +
\overline{h}\overline{e}), \hspace{10pt}  - 2(fg - bc - \overline{c}\overline{b} +
\overline{g}\overline{f}), \]
\[ - \overline{a}c + \overline{c}a + b\overline{d} - d\overline{b} +\overline{e}g - \overline{g}e - f\overline{h} + h\overline{f}, 
\]\[ c\overline{a} - a\overline{c} +\overline{d}b -\overline{b}d -g\overline{e} + e\overline{g} + \overline{f}h - \overline{h}f).\]
\eject\noindent
Again it pays to reformulate, using the spinor and vector spaces and the product rules of $\mathcal{O}(4,4)$, discussed above.  The spin space is
written as
$\mathcal{S} =
\mathcal{S}^+\oplus
\mathcal{S}^-\oplus 
\mathcal{S}^+\oplus \mathcal{S}^-$ and the elements of the vector space as triples $(s,t,\beta)$, where $(s,t) \in \R^2$ and $\beta = (u,v,x,y)\in
\V$, the eight dimensional $\mathcal{O}(4,4)$ vector space.  In this language, the $\mathcal{O}(5,5)$ spin representation is:
\[ \gamma_{s,t,\beta} = \hspace{6pt}\begin{array}{|cccc|}0&0&s&\beta\\0&0&\beta&t\\t&-\beta&0&0\\-\beta&s&0&0\end{array}\hspace{3pt}.\]
Acting on a typical element $\psi = (\alpha, \gamma, \rho, \tau)$ of $\mathcal{S}$, with $\alpha$ and $\rho$ in $\mathcal{S}^+$ and $\gamma$ and $\tau$
in
$\mathcal{S}^-$, we have:
\[ \gamma_{s,t,\beta} \hspace{3pt}\begin{array}{|c|}\alpha\\\gamma\\\rho\\\tau\end{array}\hspace{6pt} =
\hspace{6pt}\begin{array}{|c|}s\rho + \beta\tau\\t\tau + \beta\rho\\t\alpha-\beta\gamma\\s\gamma-\beta\alpha\end{array}\hspace{3pt}.\]
In this language it is easily checked that we have the required identity:
\[ \gamma^2_{s,t,\beta} = (st - \beta^2)I_\mathcal{S}.\]
Next, using the natural inner products of $\mathcal{S}^\pm$, we have a simple duality formula: putting $\phi = (\zeta, \theta, \kappa, \mu)\in
\mathcal{S} $, we have: $2\psi'\phi = \alpha.\kappa + \rho.\zeta +
\gamma.\mu +
\tau.\theta$. Specializing to the case $\phi = \psi$, we get the formula:
\[ \psi'\psi = \alpha.\rho + \gamma.\tau.\]
Using these formulas, we get:
\[ 2\psi'\gamma_{s,t,\beta}\phi = \alpha.(t\zeta-\beta\theta) + \rho.(s\kappa + \beta\mu) + \gamma.(s\theta - \beta\zeta) +
\tau.(t\mu + \beta\kappa)\]
\[ = s(\gamma.\theta+\rho.\kappa) + t(\alpha.\zeta + \tau.\mu) - \beta.(\alpha\theta +\gamma\zeta  - \rho\mu - \tau\kappa)\]
\[ = 2(\psi'\underline{\gamma}\phi).(s,t,\beta),\]
\[ \psi'\underline{\gamma}\phi = (\alpha.\zeta + \tau.\mu,\hspace{8pt} \gamma.\theta + \rho.\kappa, \hspace{8pt}\frac{1}{2}(\alpha\theta + \gamma\zeta
- \rho\mu -\tau\kappa)) = \phi'\underline{\gamma}\psi.\]
In particular we have:
\[ \psi'\underline{\gamma}\psi =  (\alpha^2 + \tau^2, \hspace{7pt}\gamma^2+\rho^2, \hspace{7pt} \alpha\gamma - \rho\tau).\]
So for the norm squared of $\psi'\underline{\gamma}\psi$, we have:
\[ |\psi'\underline{\gamma}\psi|^2 = (\alpha^2 + \tau^2)(\gamma^2+\rho^2) - (\alpha\gamma - \rho\tau)^2\]
\[ = \alpha^2\rho^2 + \gamma^2\tau^2 + 2(\alpha\gamma).(\rho\tau)\]
\[ = (\psi'\psi)^2 + \alpha^2\rho^2 - (\alpha.\rho)^2 + \gamma^2\tau^2 - (\gamma.\tau)^2 + (\alpha\gamma).(\rho\tau) -  (\rho\gamma).(\alpha\tau).\]
For the dot product between $\psi'\underline{\gamma}\psi$ and $\phi'\underline{\gamma}\psi$, we have:
\[ 2(\psi'\underline{\gamma}\psi).(\phi'\underline{\gamma}\psi) = (\gamma^2+\rho^2)(\alpha.\zeta + \tau.\mu) + (\alpha^2 + \tau^2)(\gamma.\theta +
\rho.\kappa)\]\[ - (\alpha\gamma - \rho\tau).(\alpha\theta + \gamma\zeta - \rho\mu -\tau\kappa)\]
\[ = \phi'\lambda, \]
\[ \lambda = 2(\psi'\underline{\gamma}\psi).(\underline{\gamma}\psi) = (\rho\alpha^2 + \tau(\alpha\gamma), \tau\gamma^2 + \rho(\alpha\gamma),  
\alpha\rho^2 + \gamma(\rho\tau), \gamma\tau^2 +\alpha(\rho\tau)),\]
\[ \lambda'\lambda = (\psi'\psi)|\psi'\underline{\gamma}\psi|^2.\]
Of particular importance for the following is the special case that $\psi $ is chiral: either $\rho = \tau = 0$, or $\alpha = \gamma = 0$. In the case that $\rho = \tau = 0$, we have:
\[ \psi'\underline{\gamma}\psi =  (\alpha^2, \hspace{7pt}\gamma^2, \hspace{7pt} \alpha\gamma). \]
In the case that $\alpha = \gamma = 0$, we have:
\[ \psi'\underline{\gamma}\psi = (\tau^2, \hspace{7pt}\rho^2, \hspace{7pt} - \rho\tau).\]
Then in each of these cases we have:
\[ \psi'\psi = 0, \]
\[ |\psi'\underline{\gamma}\psi|^2  = 0,\]
\[ (\psi'\underline{\gamma}\psi).(\underline{\gamma}\psi)  = 0. \]
So, in the chiral case, the vector $\psi'\underline{\gamma}\psi$ is null and orthogonal to $\underline{\gamma}\psi$.

\end{itemize}

\section{Triality axioms}
Consider a vector space $\V$, defined over a field $\F$ of zero characteristic and equipped
with the following structure:
\begin{itemize} \item A grading by a three element set $S$. \\
So $\V = \sum_{i\in S}\V_i$, where $\V_i$ is a subspace of $\V$ for each $i \in S$ and for distinct $i$ and $j$ in $S$, we have $\V_i \cap \V_j
=
\{0\}$.
\item A symmetric non-degenerate bilinear form $g: \V\times \V\rightarrow \F$.
\item A totally symmetric trilinear form $m: \V\times\V\times \V\rightarrow \F$.\end{itemize}
Denote by $\mu$ the multiplication map from $\V\times \V
\rightarrow \V$ given by the formula: $g(x, \mu(v,w)) = m(v,w,x)$,  valid for any $(v,w,x) \in \V\times \V\times \V $.  We call this structure
a triality provided that the following relations hold:
\begin{itemize}\item  For distinct $i$ and $j$ in $S$, $\V_i$ and $\V_j$ are orthogonal with respect to $g$; this entails, in particular, that
the restriction of $g$ to each $\V_i$ be non-degenerate.
\item $\mu$ is a graded multiplication: if $v\in\V_i$ and $w\in V_j$, then $\mu(v,w) \in V_k$,  where $i$, $j$ and $k$ are distinct elements of $S$.
\item The restriction of $\mu$ to $\V_i$ is trivial: for each $i$ in $S$ and for each $v\in \V_i$, $\mu(v,v) = 0$.
\item For each distinct $i$ and $j$ in $S$ and for each $v\in \V_i$ and $w \in \V_j$, we have the relation 
\[ \mu(v, \mu(v, w)) = g(v,v)w.\]
\end{itemize}
We abbreviate by writing $v^2$, $v.w$, $vw$ and $(vwx)$ for $g(v,v)$, $g(v,w)$, $\mu(v, w)$ and $m(v,w,x)$, respectively, for any $(v,w,x)\in \V\times
\V\times
\V$.  In the following $\alpha$, $\alpha'$, $\alpha''$, $\dots$, are generic elements in $\V_i$, $\beta$, $\beta'$, $\beta''$,$\dots$, are generic
elements in
$\V_j$ and $\gamma$, $\gamma'$,
$\gamma''$,$\dots$, are generic elements in
$V_k$, where
$i$,
$j$ and $k$ are the three distinct elements of $S$ (in any order). Then we have the relations:
\[ \alpha(\alpha\beta) = \alpha^2\beta, \hspace{10pt}\alpha(\alpha'\beta) + \alpha'(\alpha\beta) = 2(\alpha.\alpha')\beta, \]
\[(\alpha\hspace{-1pt}\beta).(\alpha\hspace{-1pt}\beta) \hspace{-2pt}=\hspace{-2pt} \alpha^2\hspace{-1pt}\beta^2,
\hspace{6pt}(\alpha\hspace{-1pt}\beta).(\alpha\hspace{-1pt}\beta')\hspace{-2pt} =
\hspace{-2pt}\alpha^2\hspace{-1pt}\beta.\beta',
\hspace{6pt} (\alpha\hspace{-1pt}\beta).(\alpha'\hspace{-1pt}\beta')\hspace{-3pt} +\hspace{-3pt}
(\alpha'\hspace{-1pt}\beta).(\alpha\hspace{-1pt}\beta') \hspace{-2pt}=\hspace{-2pt} 2(\alpha.\hspace{-1pt}\alpha')(\beta.\hspace{-1pt}\beta'), \]
\[ (\alpha\beta)(\alpha\gamma) = 2(\alpha\beta\gamma)\alpha - \alpha^2(\beta\gamma), \]
\[ (\alpha\beta)(\alpha'\gamma) + (\alpha\gamma)(\alpha'\beta) =  2(\alpha\beta\gamma)\alpha' +   2(\alpha'\beta\gamma)\alpha -
2\alpha.\alpha'(\beta\gamma).\]
\eject\noindent\section{The Jordan algebra associated to a triality} We next introduce the vector space $\J  = \V\oplus \F\oplus \F\oplus \F$.  A
typical element
$J$ of
$\J$ is written as a three by three symmetric matrix:
\[  J = \hspace{5pt}\begin{array}{|ccc|}a&\gamma&\beta\\\gamma&b&\alpha\\\beta&\alpha&c\end{array}.\]
Here $a$, $b$ and $c$ are in $\F$ and $\alpha$, $\beta$ and $\gamma$ lie in $\V_i$, $\V_j$ and $\V_k$, where $S = \{i,j,k\}$.
The "matrix product" $JJ'$ of two such elements $J$ and $J'$ makes sense (although it is not in general symmetric, so it need not lie in $\J$):
\[ JJ' = \hspace{5pt} \begin{array}{|ccc|}a&\gamma&\beta\\\gamma&b&\alpha\\\beta&\alpha&c\end{array}\hspace{5pt}
\begin{array}{|ccc|}a'&\gamma'&\beta'\\\gamma'&b'&\alpha'\\\beta'&\alpha'&c'\end{array}\hspace{5pt} = \hspace{5pt}
\begin{array}{|ccc|}aa' + \gamma.\gamma' + \beta.\beta'&a\gamma' + b'\gamma +\beta\alpha'&a\beta' + c'\beta + \gamma\alpha'\\b\gamma' + a'\gamma +
\alpha\beta'&bb' +
\alpha.\alpha' +
\gamma.\gamma'&b\alpha' + c'\alpha + \gamma\beta'\\c\beta'+  a'\beta + \alpha\gamma'&c\alpha'+b'\alpha + \beta\gamma'&cc' +
\beta.\beta' +
\alpha.\alpha'\end{array}.\]
Then the Jordan product is $J\circ J' = \frac{1}{2}(JJ' + J'J)$, which is symmetric and lies in the space $\J$.  Note that the structure of this
product is invariant under a permutation of the elements of $S$ accompanied by a corresponding permutation of the triple $\{a,b,c\}$. In particular we
have the Jordan square:
\[ J^2 = J\circ J = \hspace{5pt}
\begin{array}{|ccc|}a^2 + \gamma^2 + \beta^2&(a+b)\gamma +\alpha\beta&(c+a)\beta + \gamma\alpha\\(a+ b)\gamma +
\alpha\beta&b^2 +
\alpha^2 +
\gamma^2&(b+c)\alpha + \beta\gamma\\(c+a)\beta+ \gamma\alpha&(b + c)\alpha+ \beta\gamma&c^2 +
\beta^2 +
\alpha^2\end{array}.\]
The trace of the Jordan product gives $\J$ a non-degenerate metric, denoted $G$ and also written as a dot product:
\[ G(J, J') = J.J' = \tr(JJ') = \tr(J\circ J') = aa' + bb' + cc' + 2\alpha.\alpha' + 2\beta.\beta' + 2\gamma.\gamma',\]
\[ G(J, J) = J.J = \tr(J^2) = a^2 + b^2 + c^2 + 2\alpha^2 + 2\beta^2 + 2\gamma^2.\]
The determinant and adjoint matrix of $J$, $\det(J)$ and $\textrm{ad}(J)$ respectively, have the natural definitions:
\[ \det(J) = abc - a\alpha^2 - b\beta^2 - c\gamma^2 + 2(\alpha\beta\gamma), \]
\[ \textrm{ad}(J) = \hspace{5pt}\begin{array}{|ccc|}bc - \alpha^2&\alpha\beta-c\gamma&\gamma\alpha - b\beta\\\alpha\beta - c\gamma&ca -
\beta^2&\beta\gamma - a\alpha\\\gamma\alpha - b\beta&\beta\gamma-a\alpha&ab -
\gamma^2\end{array}.\]
Note that twice the trace of the adjoint gives $J$ another non-degenerate metric, denoted $H$: $H(J, J) = 2\tr(\textrm{ad}(J)) = 2bc + 2ca + 2ab - 2\alpha^2 -
2\beta^2 - 2\gamma^2$.\eject\noindent We then have the following key properties (where $I$ is the three by three identity matrix):
\[ G(H, H) + G(J, J) = (\tr(J))^2.\] 
\[ J\textrm{ad}(J) =  \textrm{ad}(J)J  = \frac{1}{2}J\circ\textrm{ad}(J) = \det(J)I, \]
\[ J.\textrm{ad}(J) = 3\det(J),\]
\[ \textrm{ad}(\textrm{ad}(J)) = \det(J)J. \]
The first and second identities are proved directly.  The third identity follows from the second by taking the trace.  For the fourth
identity, using the symmetry the structure under the cyclic permutation of the elements of $S$, it suffices to check the relation for one diagonal and
one off-diagonal entry.  We choose the
$(11)$ and $(12)$ matrix elements:
\[  [\textrm{ad}( \textrm{ad}(J))]_{11} =  \textrm{ad}(J)_{22}\textrm{ad}(J)_{33} -   \textrm{ad}(J)_{23}^2\]
\[ = (ac - \beta^2)(ab - \gamma^2) - (a\alpha - \beta\gamma)^2\]
\[ = a(abc - b\beta^2 - c\gamma^2 - a\alpha^2 + 2(\alpha\beta\gamma)) + \beta^2\gamma^2 - (\beta\gamma)^2\]
\[ = a\det(J) = J_{11}\det(J).\]
\[ [\textrm{ad}(\textrm{ad}(J))]_{12} =  \textrm{ad}(J)_{13}\textrm{ad}(J)_{32} -   \textrm{ad}(J)_{12}\textrm{ad}(J)_{33}\]
\[ = (\gamma\alpha - b\beta)(\beta\gamma - a\alpha) - (\alpha\beta- c\gamma)(ab - \gamma^2)\]
\[ = (\alpha\beta)(\alpha\gamma) + \gamma^2\alpha\beta \gamma^2 - b\beta^2\gamma- a\alpha^2\gamma + abc\gamma - c\gamma^2\gamma\]
\[ = 2(\alpha\beta\gamma)\gamma - b\beta^2\gamma- a\alpha^2\gamma + abc\gamma - c\gamma^2\gamma  = \gamma\det(J) = J_{12}\det(J).\]
The matrix $\textrm{ad}(J)$ is a homogeneous quadratic in the elements of $J$, so there exists a unique bilinear commutative product, denoted $J\times
J'$, called the cross-product, such that
$J\times J =
2\textrm{ad}(J)$. Specifically we have:
\[ J\times J' = \hspace{5pt}\begin{array}{|ccc|}bc' + cb' - 2\alpha.\alpha'&\alpha\beta' + \beta\alpha'-c\gamma' - c'\gamma&\gamma\alpha' + \alpha\gamma' -
b\beta' - b'\beta\\\alpha\beta' +\beta\alpha' - c\gamma' - c'\gamma&ca' + ac' -
2\beta.\beta'&\beta\gamma' + \gamma\beta' - a\alpha' - a'\alpha\\\gamma\alpha' + \alpha\gamma' - b\beta' - b'\beta&\beta\gamma' + \gamma\beta'-a\alpha' -
a' \alpha&ab' + ba' -
2\gamma.\gamma'\end{array}.\]
Similarly, the quantity  $\det(J)$ is a homogeneous cubic in the elements of $J$, so there exists a unique totally symmetric trilinear form, denoted
$(J, J', J'')$, such that
$(J, J, J) = 6\det(J)$. Specifically we have:
\[ (J, J', J'') = bc'a''+ cb'a'' + ca'b'' + ac'b'' + ab'c'' + ba'c'' - 2a\alpha'.\alpha''\]\[ - 2a''\alpha.\alpha'- 2a'\alpha''.\alpha - 2b\beta'.\beta'' -
2b''\beta.\beta'- 2b'\beta''.\beta - 2c\gamma'.\gamma''- 2c'\gamma''.\gamma - 2c''\gamma.\gamma' \]\[+ 2(\beta\gamma'\alpha'') + 2(\gamma\beta'\alpha'') +
2(\gamma\alpha'\beta'') + 2(\alpha\gamma'\beta'') + 2(\alpha\beta'\gamma'') + 2(\beta\alpha'\gamma'').\]
We notice, by direct calculation, the following important relation:
\[ (J,J',J'') = (J\times J').J''.\]
In the language of the cross product, the relation $\textrm{ad}(\textrm{ad}(J)) = \det(J)J$ now reads:
\[ 3(J\times J) \times (J\times J) = 4(J,J,J)J.\]
Fully polarizing this identity and then specializing gives the following relations:
\[ (J\times K)\times(L\times M) + (J\times M)\times(K\times L) + (J\times L)\times(M\times K) \]\[= J(K, L, M) + K(J, L, M) + L(J, M, K) + M(J, K, L).\]
\[ 2(J\times K)\times(J\times L) + (J\times J)\times(K\times L) = 2J(J, K, L) + K(J, J, L) + L(J, J, K).\]
\[ 2(J\times K)\times(J\times K) + (J\times J)\times(K\times K) = 2J(J, K, K) + 2K(K,J, J).\]
\[ 3(J\times J)\times(J\times K) = 3J(J, J, K) + K(J, J, J).\]
Dotting the last two of these relations with $J\times K$ and $K\times K$, respectively, gives the relations:
\[ 4\det(J\times K) + 4\det(J)\det(K) = (J,K,K)(J,J,K). \]
\[ (J\times J, J\times K, K\times K) = 12\det(J)\det(K) + (J,J,K)(J,K,K).\]
\[ \det(J\times J) = 8 \det(J)^2.\]
\eject\noindent
Finally we obtain a trickier trilinear identity:
\[ J\times(J\times J) + \tr(J)J\times J + \tr(J\times J)J + (2\det(J) - \tr(J)\tr(J\times J))I = 0.\]
One way to derive this identity is to specialize the identity for $(J\times J)\times (J\times K)$ to the case that $K = -I$, using the fact (easily verified)
that
$(-I)\times J = J - \tr(J)I$.  Alternatively, by the symmetry of $S$, it suffices to check this identity for one diagonal and one off-diagonal entry: 
we choose the
$(11)$ and $(12)$ matrix elements.  The $(12)$ entry of the left-hand side is:
\[  (J\times(J\times J))_{12} + \tr(J)(J\times J)_{12} + \tr(J\times J)J_{12}\]
\[ =  J_{23}(J\times J)_{13} \hspace{-2pt}+\hspace{-2pt} J_{13}(J\times J)_{23}\hspace{-2pt} -\hspace{-2pt}J_{33}(J\times J)_{12}\hspace{-2pt}
-\hspace{-2pt} J_{12}(J\times J)_{33} \hspace{-2pt}+\hspace{-2pt} 2\tr(J)(\alpha\beta-c\gamma)\hspace{-2pt} +\hspace{-2pt}
\tr(J\times J)\gamma\]
\[ = 2\alpha(\gamma\alpha\hspace{-2pt} -\hspace{-2pt} b\beta)\hspace{-2pt} + \hspace{-2pt}2\beta(\beta\gamma\hspace{-2pt} -\hspace{-2pt}
a\alpha)\hspace{-2pt} -\hspace{-2pt} 2c(\alpha\beta\hspace{-2pt} -\hspace{-2pt} c\gamma)\hspace{-2pt} -\hspace{-2pt} 2\gamma(ab\hspace{-2pt}
-\hspace{-2pt}\gamma^2) \hspace{-2pt}+ \hspace{-2pt}2(a \hspace{-2pt}+\hspace{-2pt} b\hspace{-2pt} +\hspace{-2pt}
c)(\alpha\beta\hspace{-2pt}-\hspace{-2pt}c\gamma)\hspace{-2pt} +\hspace{-2pt}
\tr(J\times J)\gamma\]
\[ = \gamma(2\alpha^2 + 2\beta^2 + 2\gamma^2 - 2ab - 2ca - 2bc + \tr(J\times J)) = 0.\]
 The $(11)$ entry of the left-hand side is:
\[  (J\times(J\times J))_{11} + \tr(J)(J\times J)_{11} + \tr(J\times J)J_{11} +  2\det(J) - \tr(J)\tr(J\times J)\]
\[ =  J_{22}(J\hspace{-2pt}\times\hspace{-2pt} J)_{33}\hspace{-2pt} + \hspace{-2pt}J_{33}(J\hspace{-2pt}\times \hspace{-2pt}J)_{22} \hspace{-2pt}-
\hspace{-2pt}2J_{23}.(J\hspace{-2pt}\times\hspace{-2pt} J)_{23}\hspace{-2pt}  +\hspace{-2pt}
\tr(J)(2bc\hspace{-2pt} -\hspace{-2pt} 2\alpha^2) \hspace{-2pt}- \hspace{-2pt}(b\hspace{-2pt}+\hspace{-2pt}c)\tr(J\hspace{-2pt}\times\hspace{-2pt}
J)\hspace{-2pt} +\hspace{-2pt} 2\det(J)\]
\[ =  2b(ab \hspace{-2pt}-\hspace{-2pt} \gamma^2)\hspace{-2pt} +\hspace{-2pt} 2c(ca\hspace{-2pt} -\hspace{-2pt} \beta^2)\hspace{-2pt} -\hspace{-2pt}
4\alpha.(\beta\gamma\hspace{-2pt} -\hspace{-2pt} a\alpha)\hspace{-2pt} +\hspace{-2pt} (a\hspace{-2pt} +\hspace{-2pt} b\hspace{-2pt} +\hspace{-2pt}
c)(2bc\hspace{-2pt} -\hspace{-2pt} 2\alpha^2)\hspace{-2pt} -\hspace{-2pt}
(b\hspace{-2pt}+\hspace{-2pt}c)\tr(J\hspace{-2pt}\times\hspace{-2pt}J)\hspace{-2pt} +\hspace{-2pt} 2\det(J)\]
\[ =  2ab^2+2ac^2 + 2b^2c + 2c^2b - 2b\gamma^2 - 2c\beta^2 + 2abc - 4(\alpha\beta\gamma) + 2a\alpha^2 \]\[ -2(b + c)(ab + bc + ca -
\beta^2 - \gamma^2) + 2\det(J)\]
\[ =  2(a\alpha^2 + b\beta^2 + c\gamma^2 - abc - 2(\alpha\beta\gamma)  + \det(J)) = 0.\]
This completes the proof of the identity. Polarizing the identity and specializing appropriately gives the following relations:
\[ 0 = J\times(K\times L) +  K\times(L\times J) +  L\times(J\times K) \]\[ + \tr(J)K\times L + \tr(K)L\times J +  \tr(L)J\times K + \tr(K\times L)J  +
\tr(L\times J)K + \tr(J\times K)L\]\[ + ((J, K, L) - \tr(J)\tr(K\times L) - \tr(K)\tr(L\times J) - 
\tr(L)\tr(J\times K))I.\]
\[ 0 = 2J\hspace{-2pt}\times\hspace{-2pt}(J\hspace{-2pt}\times\hspace{-2pt} K)\hspace{-2pt} + \hspace{-2pt}
K\hspace{-2pt}\times\hspace{-2pt}(J\hspace{-2pt}\times\hspace{-2pt}J) \hspace{-2pt} +\hspace{-2pt} 2\tr(J)J\hspace{-2pt}\times\hspace{-2pt} K
\hspace{-2pt}+\hspace{-2pt}
\tr(K)J\hspace{-2pt}\times\hspace{-2pt} J\hspace{-2pt} +\hspace{-2pt} 2\tr(J\hspace{-2pt}\times\hspace{-2pt} K)J\hspace{-2pt}  +\hspace{-2pt}
\tr(J\hspace{-2pt}\times\hspace{-2pt} J)K \]\[ + ((J, J, K) - 2\tr(J)\tr(J\hspace{0pt}\times\hspace{0pt} K) \hspace{0pt}-
\hspace{0pt}\tr(K)\tr(J\times J))I.\]
\section{The condition $J\times J = 0$}
Consider the equation $J\times J = 0$.  Written out, this equation gives the following six conditions:
\[ \alpha^2 = bc, \hspace{10pt} \beta^2 = ca, \hspace{10pt} \gamma^2 = ab, \hspace{10pt}\beta\gamma = a\alpha, \hspace{10pt}\gamma\alpha = b \beta,
\hspace{10pt}\alpha\beta= c\gamma.\] Clearly $J = 0$ solves this equation, so henceforth we assume that $J \ne 0$.  After using the symmetry of
$S$, we may reduce to the analysis of two cases: first that $a = b = c = 0$ and $\beta \ne 0$ and second that $b \ne 0$. \\\\ In the following we
repeatedly use the fact that for a given non-zero null vector $\beta$, the equation $\beta\alpha = 0$ has the general solution $\alpha = \beta\gamma$,
for some $\gamma$. Indeed we have $\gamma = \tilde{\beta}\alpha$, where $\tilde{\beta}$ is any vector normalized against $\beta$ by the relation
$2\beta.\tilde{\beta} = 1$. Then
$\gamma$ is determined uniquely up to a transformation of the form
$\gamma\rightarrow
\gamma +
\beta\alpha'$, where
$\alpha'$ is arbitrary.  Further if $n$ is the dimension of $\V$, the vector space formed by the solutions $\alpha$ of the equation $\beta\alpha = 0$,
for a given non-zero null vector 
$\beta$ has dimension $\frac{n}{2}$.  More generally the space of solutions of the equation $\beta\alpha = 0$ is either $(i)$ $0$-dimensional when
$\beta^2 \ne 0$, $(ii)$ $\frac{n}{2}$-dimensional when $\beta^2 = 0$ and $\beta\ne 0$, or $(iii)$ $n$-dimensional when $\beta = 0$.  Note that case
$(ii)$ cannot occur in the octavic case.  For us
$n = 8$.
\begin{itemize}\item In the first case, with
$a  = b = c = 0$ and
$\beta
\ne 0$, the equations become:
\[ a = b = c = 0 = \alpha^2 = \beta^2 = \gamma^2 = 0, \hspace{7pt}\alpha\beta = 0, \hspace{7pt}\beta\gamma = 0,
\hspace{7pt}\gamma\alpha = 0,\hspace{10pt}\beta\ne 0.\]
\[ J = \begin{array}{|ccc|}0&\gamma&\beta\\\gamma&0&\alpha\\\beta&\alpha&0\end{array}\hspace{3pt}, \hspace{20pt} J^2 = 0.\]
Since $\beta \ne 0$, we may solve for $\alpha$ and $\gamma$ as follows:
\[ \alpha = \beta\tilde{\gamma}, \hspace{10pt}\gamma = \beta\tilde{\alpha}.\]
Then all equations are satisfied identically except for the equation $0 = \alpha\gamma$, which, since $\beta^2 =0$, gives the equation $0 =
2(\beta\tilde{\alpha}\tilde{\gamma})\beta$, so we have the additional scalar constraint: $(\beta\tilde{\alpha}\tilde{\gamma}) = 0$.  The
freedom in the solution, given $\beta$ is
$\tilde{\alpha}\rightarrow
\tilde{\alpha} +
\beta \gamma'$ and   $\tilde{\gamma}\rightarrow \tilde{\gamma} +
\beta \alpha'$, where $\alpha'$ and $\gamma'$ are arbitrary.  The number of degrees of freedom in the solution is represented by $n -1$ for the null
vector $\beta$, $\frac{n}{2}$ each for $\tilde{\alpha}$ and $\tilde{\gamma}$ less one for the scalar constraint, giving $2n -2$ degrees of freedom. 
Projectively the freedom here is $2n - 3$.  For $n = 8$ this is a thirteen dimensional manifold.  
\item In the second case, with $b\ne 0$, we need:
\[ c = \frac{\alpha^2}{b}, \hspace{10pt} a = \frac{\gamma^2}{b}, \hspace{10pt}\beta = \frac{\alpha\gamma}{b}, \hspace{10pt}\beta^2 = ca,
\hspace{10pt} \beta\gamma = a\alpha, \hspace{10pt}\alpha\beta= c\gamma.\] 
These equations may be regarded as specifying $a$, $c$ and $\beta$, given
$\alpha$, $\gamma$ and $b$ and then all the equations are satisfied.  $J$ is determined completely, with $\alpha$ and $\gamma$ and $b\ne 0$ being
arbitrary (giving
$2n$ projective degrees of freedom):
\[ J = \begin{array}{|ccc|}\frac{\gamma^2}{b}&\gamma&\frac{\alpha\gamma}{b}\\\gamma&b&\alpha\\
\frac{\alpha\gamma}{b}&\alpha&\frac{\alpha^2}{b}\end{array},
\hspace{20pt}J^2 = \frac{1}{b}(\alpha^2 +
\gamma^2 + b^2)J.\] 
\end{itemize}
Thus the structure of the projective space of all $J$, such that $J\times J = 0$, where $J$ is non-zero and $J\equiv J'$  iff $J = tJ'$ with $t$ a
non-zero real number, is that it is covered by six sets: 
\begin{itemize}\item For each $i \in S$, a space $C_i$, an appropriate permutation of the projective space of all triples $(\beta\tilde{\gamma}, \beta,
\beta\tilde{\alpha})$, with
$(\beta\tilde{\alpha}\tilde{\gamma}) = 0$ and $\beta \ne 0$. 
\item For each $i \in S$, $U_i$, the linear span of  $\V_j$ and $\V_k$ in $\V$, where $S = \{i, j,k\}$.
\end{itemize}
Note that $C_0 = C_i\cup C_j\cup C_k$ is a closed set, of projective dimension $2n-3$ and disjoint from the union of the three open sets $\{U_i: i\in
S\}$, which each have projective dimension $2n$.  In the octavic case $C_0$ is empty.  Then the space is the Moufang octavic projective
plane.  The three spaces $U_i$ then form three affine spaces of type $\R^{16}$, which together patch the plane.
\\\\In the octavic case, Baez describes each solution of the equation $J\times J = 0$, up to a non-zero real scale
factor, as an outer product, as follows \cite{Ba1}:
\[ J =  \hspace{7pt}\begin{array}{|c|}\alpha\\\beta\\\gamma\end{array}\hspace{7pt}\begin{array}{|ccc|}\alpha&\beta&\gamma\end{array}
\hspace{7pt} =
\hspace{7pt}\begin{array}{|ccc|}\alpha^2&\alpha\beta&\alpha\gamma\\\alpha\beta&\beta^2&\beta\gamma\\\alpha\gamma&\beta\gamma&\gamma^2\end{array}\hspace{7pt}.\]
Here the three equations $(\alpha\beta\gamma)\alpha = \alpha^2\beta\gamma$, $(\alpha\beta\gamma)\beta = \beta^2\gamma\alpha$ and
$(\alpha\beta\gamma)\gamma = \gamma^2\alpha\beta$ must all hold.  If at least one of the three vectors $\alpha$, $\beta$ or $\gamma$ vanish these
conditions hold automatically. On the other hand, if each of the three vectors is non-zero, then these conditions amount to the requirement that
$\gamma$ be a non-zero multiple of
$\alpha\beta$.  However, in the case of split octaves, it can be shown that not all non-zero solutions of the equation $J\times J = 0$ can be
parametrized as an outer product in this way.  A seemingly better approach (valid in both the octavic and split octavic cases) is to
write
$J$ in the form
$K\times K$, where
$\det(K) = 0$.  

 \eject\noindent
\section{The projective geometry of the Jordan algebra: Points, lines, "fat points" and "fat lines"}
In the octavic case, associated to the Jordan algebra is a projective geometry, called the (real) Moufang plane.  The geometry consist of points
and lines, together with an incidence relation:
\begin{itemize}\item A point is an equivalence class under scaling by non-zero reals of elements $J\ne 0$ of the Jordan algebra, such that $J\times J
= 0$
\item  A line is an equivalence class under scaling by non-zero reals of elements $K\ne 0$ of the dual Jordan algebra, such that $K\times K
= 0$
\item The point $J$ is incident on the line $K$ iff $J = M\times K$, for some $M$ in the dual algebra such that $M\times M = 0$ iff $K = L\times J$,
for some $L$ in the algebra, such that $L\times L = 0$.  Intuitively, we think of $J$ as the (unique) intersection of the lines $M$ and $K$, whereas
$K$ is the (unique) line though $J$ and $L$.
\end{itemize}
Note that when the point $J$ is on the line $K$, the relation $J.K = 0$ holds (since $J.K = J.(L\times J) = (J\times J).L = 0$), but the converse is
\emph{not} true (despite a statement to that effect in Freudenthal, transmitted in Baez).  The condition that $J.K$ should vanish is just one real
condition and is not enough to give the desired incidence.\\\\The structure of the Moufang plane depends crucially on the positivity properties of the
octavic norm-squared, so does not go through either in the case of the split-octaves or in the case of the complexified octaves.  Instead we are
led to introduce a new kind of geometric object called a "fat point" (together with its dual, the "fat line").
\begin{itemize}\item A point is an equivalence class under scaling by non-zero reals of elements $J\ne 0$ of the Jordan algebra, such that
$J\times J = 0$.
\item  A line is an equivalence class under scaling by non-zero reals of elements $K\ne 0$ of the dual Jordan algebra, such that $K\times K
= 0$.
\item The point $J$ is incident on the line $K$ iff $J = M\times K$, for some $M$ in the dual algebra such that $M\times K = 0$ iff $K = L\times J$,
for some $L$ in the algebra, such that $L\times L = 0$.  Intuitively we think of $J$ as the (unique) intersection of the lines $M$ and $K$, whereas
$K$ is the (unique) line though $J$ and $L$.
\item A "fat point" is a pair $(J, J')$ in the Jordan algebra, such that $J$ and $J'$ are linearly independent and yet $J\times J = J\times
J' = J'\times J' =0$.
\item A "fat line" is a pair $(K, K')$ in the dual Jordan algebra, such that $K$ and $K'$ are linearly independent and yet $K\times K =
K\times K' = K'\times K' =0$.
\end{itemize}
Note that in all cases, if $J\times J = J'\times J' = 0$ and if we put $K = J\times J'$, then by our Jordan identities, we always have
$K\times K = 0$.\\\\
In the octavic case, if $J\times J = J'\times J' = 0$ and $J$ and $J'$ are
linearly independent, then the cross-product $J\times J'$ is never zero so the "fat points" are never needed.  Here, we regard
distinct points
$J$ and
$J'$ as defining a line in the case that $J\times J' \ne  0$ and a fat point in the case that $J\times J' = 0$.
\\\\
There is then a seemingly natural generalization: a generalized point can be considered to be a polynomial $J(\underline{x})$ in some real variables
$\underline{x}$, such that $J(\underline{x})\times J(\underline{x}) = 0$, with a dual idea for a generalized line.  A particularly nice example is the
quadratic point $J(s, t) = s^2J + 2stJ' + t^2J''$, where $s$ and $t$ are free real parameters.  Then the condition $J(s,t)\times J(s,t) = 0$, valid for
all $s$ and $t$, amounts to the five conditions on the coefficients:
$J\times J = 0$, 
$J\times J' = 0$,  $J\times J'' + 2J'\times J' = 0$,  $J'\times J'' = 0$ and  $J''\times J'' = 0$.  Note that the "fat point" can be
regarded as a special case of the quadratic point, where either $J' = 0$, or $J'' = 0$. \\\\For an ordinary point it is natural to seek a
parametrization of the form
$J = K\times K$, where
$\det(K) = 0$.  More generally for a generalized point, we can seek an analogous formula: $J(\underline{x}) = K(\underline{x})\times
K(\underline{x})$, where
$K(\underline{x})$ is a curve lying in the manifold
$\det(K) = 0$.  Here we are always exploiting the relation $(K\times K)\times (K\times K)  = 8\det(K)K$ which forces $J\times J = 0$, for $J =
K\times K$, whenever $\det(K) = 0$.  So for example in the case of the quadratic point, we would look for an expression
$J(s,t) = K(s,t)\times K(s,t)$, where
$K(s,t)$ is linear in $s$ and $t$.  Written out, the equations are $J = K \times K$, $J' = K\times K'$ and  $J'' = K'\times K'$, where
$K(s,t) = sK + tK'$.  Further generalizations can be contemplated, allowing, for example, rational or analytic curves $J(\underline{x})$ and
parametrizations
$J(\underline{x}) = K(\underline{x})\times K(\underline{x})$, with $K(\underline{x})$ also rational or analytic, respectively.     

\eject\noindent
\section{The solvability of equation $K\times K = J$}
Consider the equation $K\times K = J$, where $J$ is given and we wish to find $K$.  If we take the cross-product of this equation with itself, we get
the equation:
\[ J\times J = (K\times K)\times (K\times K) = 8\det(K)K.\]
Also we have the formula:
\[ \det(J) = \det(K\times K)  = 8\det(K)^2.\]
So the problem splits into three:
\begin{itemize}\item First we assume that $\det(J) < 0$.  Then the last equation shows that there are no solutions for $K$.
\item Second we assume that $\det(J) > 0$.  Then the last equation shows that $\det(K) \ne 0$.  By replacing $K$ by $-K$ (thus changing
the sign of the determinant of $K$), if necessary, we may assume that $\det(K) > 0$.  Then we have the solution:
\[ K = \frac{1}{\sqrt{8\det(J)}}(J\times J).\]
Thus in the case $\det(J) > 0$ there are exactly two solutions for $K$:
\[ K = \pm \frac{1}{\sqrt{8\det(J)}}(J\times J).\]
More generally, in the case that $\det(J) \ne 0$, then the formula:
\[ K = \pm \frac{1}{\sqrt{8\epsilon \det(J)}}(J\times J),\]
is the general solution of the equation $K\times K = \epsilon J$, where $\epsilon$ is the sign of $\det(J)$.   It is easily checked that this
formula for $K$ does solve the given equation, as required.
\end{itemize}
It remains to treat the case that $\det(J) = 0$.  Since $\det(J) = 8\det(K)^2$, we need also $\det(K) = 0$.  Note that conversely, if $K\times K =
J$ and
$\det(K) = 0$, then $\det(J) = 0$.  Further, once we have $\det(K) = 0$, we also have $J \times J = 8\det(K)K = 0$, so there are no solutions, unless
$J\times J = 0$. Also if $J\times J = 0$, then it follows that $\det(J) = 0$. 
\eject\noindent
So we are reduced to studying the following problem: solve
$K\times K = J$, for $K$ given
$J$, where $J$ obeys the equation $J\times J = 0$.  We prove solvability for this case, as follows.  Clearly if $J = 0$, then a solution is $K = 0$, so
we can assume that $J\ne 0$.  First recall the identity obeyed by the cross-product, proved earlier:
\[ (J\times L)\times (J\times L) = -\frac{1}{2}((J\times J)\times (L\times L)) + J(L,L,J) + L(J,J,L). \]
Specializing this relation to the case that $J\times J = 0$ gives the equation:
\[  (J\times L)\times (J\times L) =  J(L,L,J).\]
Now put $L = P\times P$, for some $P$.  Then $L\times L = 8\det(P)P$ and $(L,L,J) = (L\times L).J = 8\det(P)P.J$.  So putting $K = sJ\times
(P\times P)$, for some scalar $s$, we have: $K\times K = (8 s^2 \det(P)P.J)J$.  If we can always choose $P$ such that $\det(P)P.J > 0$, then we can
choose $s$ so that $K\times K = J$ and we are done.  So, for example, if $\tr(J) > 0$, we may take $P = I$ and we are done.  More generally, if $J$ has
at least one non-zero diagonal element, then it is easy to find a diagonal $P$ which works.  \\\\So the only undecided case is the case that each
diagonal entry of
$J$ is zero (and therefore at least one non-diagonal entry is non-zero).  Then find a Jordan matrix $Q$ also with all its diagonal entries zero, such
that
$Q.J >  0$.  We can do this even with just one appropriately chosen non-zero off-diagonal entry for $Q$.  Then put $P(t) = tI + Q$.  Then we have the
characteristic polynomial formula:
\[ \lim_{t\rightarrow \infty} t^{-3}\det(P(t))(P(t).J) = Q.J > 0.\]
The point here is that only the $t^3$ term in the characteristic polynomial survives in the limit.
So for large enough positive real
$t$, we have  $\det(P(t))(P(t).J)> 0$ and we are done. \\
\\
Summarizing we have proved that the equation $K\times K = J$ is solvable for $K$ given $J$ when  $\det(J) > 0$ (with exactly two solutions) and when
$J\times J = 0$ and is otherwise not solvable; also the equation $tK\times K = J$, is solvable for $K$ and the scalar $t$, except in the case that
$\det(J) = 0$ and
$J\times J\ne 0$. 
\eject\noindent
\section{Solving the equation $K\cross K = 2J$ with $J\cross J = 0$}
We wish to solve explicitly the equation $K\cross K = 2J$, where $J$ is given and $J\cross J = 0$, keeping track of the number of degrees of freedom
in the solution, in each case.  We may use the permutation symmetry of the Jordan algebra to reduce to two cases: first where $J$ has its middle
diagonal entry non-zero and second when all three diagonal entries are zero.  
\begin{itemize}\item We first assume that $J$ has at least one non-zero diagonal element.  After an appropriate permutation we may write $J$, $K$ and
$\frac{1}{2}K\times K$ as follows: 
\[ J = 
\begin {array}{|ccc|}
 \frac{\tau^2}{s}&\tau&\frac{\rho\tau}{s}
\\ \tau &s&\rho
\\ \frac{\rho\tau}{s}&\rho&\frac{\rho^2}{s}
\end {array}\]
\[K= \left|
\begin {array}{ccc}
 a&\gamma&\beta
\\ \gamma&b&\alpha
\\ \beta&\alpha&c
\end {array}
\right|\]
\[\frac{1}{2} K\cross K = 
\begin {array}{|ccc|}
 \ bc-\alpha^2&\alpha\beta-c\gamma&\alpha\gamma-b\beta
\\ \alpha\beta-c\gamma&ac-\beta^2&\beta\gamma-a\alpha
\\ \alpha\gamma-b\beta&\beta\gamma-a\alpha&ab-\gamma^2
\end {array}\]
Here $s$ is a non-zero real number.  If now $K\times K = 2J$, and $J\times J = 0$, we have the auxiliary relation:
\[0= 4J \cross J = (K \cross K)\cross(K \cross K)= 8\det(K) K. \]
Hence $\det(K) = 0$, giving the equation:
\[2(\alpha\beta\gamma)=a\alpha^2+b\beta^2+c\gamma^2-abc\]
We need to solve the following equations:
\[\rho^2=s(ab-\gamma^2),\hspace{10pt}s=ac-\beta^2,\hspace{10pt}\tau^2=s(bc-\alpha^2)\]
\[ \rho=\beta\gamma-a\alpha,\hspace{10pt}\rho\tau=s(\alpha\gamma-b\beta),\hspace{10pt}\tau =\alpha\beta-c\gamma.\]
By direct computation, using the determinant formula for $(\alpha\beta\gamma)$ and the relation $ac - \beta^2 = s$, we see that three of these
equations are identically satisfied:
\[\tau^2=(\alpha\beta-c\gamma)^2=\alpha^2\beta^2-2c(\alpha\beta\gamma)+c^2\gamma^2=s(bc-\alpha^2),\]
\[\rho^2=(\beta\gamma-a\alpha)^2=\beta^2\gamma^2-2a(\alpha\beta\gamma)+a^2\alpha^2= s(ab-\gamma^2),\]
\[\rho\tau=(\alpha\beta-c\gamma)(\beta\gamma-a\alpha)=2(\alpha\beta\gamma)\beta-\beta^2\alpha\gamma-a\alpha^2\beta-c\gamma^2\beta+ac\alpha\gamma=s(\alpha\gamma-b\beta).\]
This reduces the problem to that of finding the solutions of the equations:
\[ s=ac-\beta^2, \hspace{10pt}2(\alpha\beta\gamma)=a\alpha^2+b\beta^2+c\gamma^2-abc, \]
\[\rho=\beta\gamma-a\alpha, \hspace{10pt} \tau =\alpha\beta-c\gamma.\]
To find the general solution to these equations, we construct $\gamma$ first.  Multiplying the equation $\rho = \beta \gamma - a\alpha$ on both sides by
$\beta$ gives:
\[ \rho\beta= \beta(\beta \gamma - a\alpha) = \beta^2\gamma-a\alpha\beta \]\[ 
= \beta^2\gamma-a(\tau +c\gamma)  = -s\gamma - a\tau.\]
Since $s \ne 0$, we may solve this equation for $\gamma$:
\[\gamma=- \frac{1}{s}(\rho\beta+a\tau).\]
Similarly, we obtain a formula for $\alpha$:
\[(s-ac)\alpha=-\beta^2\alpha =-\beta^2\alpha-c\beta\gamma+c\beta\gamma\]
\[\alpha =-\frac{1}{s}(\beta(\alpha\beta-c\gamma)+c(\beta\gamma-a\alpha))=-\frac{1}{s}(\beta\tau+c\rho).\]
\eject\noindent
Finally, we use the determinant condition to determine $b$:
\[ 0 = - 2(\alpha\beta\gamma)+ a\alpha^2+b\beta^2+c\gamma^2 -abc\]
\[= - \frac{2}{s}\beta.(\rho\tau + bs\beta) + a(bc-\frac{1}{s}\tau^2) - bs + c(ab-\frac{1}{s}\rho^2) \]\[=-\frac{c}{s}\rho^2-\frac{a}{s}\tau^2 -
\frac{2}{s}(\beta\rho\tau) - 2b(\beta^2 - ac) - bs,\]
\[b= \frac{1}{s^2}(c\rho^2+2(\beta\rho\tau)+a\tau^2).\]
Summarizing the general solution is given by the formulas:
\[ \alpha = -\frac{1}{s}(\beta\tau+c\rho), \]
\[\gamma=- \frac{1}{s}(\rho\beta+a\tau), \]
\[ b = \frac{1}{s^2}(c\rho^2+2(\beta\rho\tau)+a\tau^2).\]
Here $\beta$, $a$ and $c$ are subject only to the condition:
\[ \beta^2 - ac = - s.\]
So there are $n + 1$ degrees of freedom in the solution.
\eject\noindent\item  Henceforth we may assume that the diagonal elements of $J$ are all zero.  If $J = 0$, then section five above shows how to obtain and
parametrize all the solutions $K$, such that $K\times K = 2J = 0$, so henceforth we assume that $J\ne 0$.  We first solve for
$K$ under the additional assumption that $K$ has a non-zero diagonal entry,  which we may take to be $c$.  By the permutation  symmetry, the one remaining
case is the case that all diagonal entries of $K$ are zero.  With $c \ne 0$,  the matrix equation to be solved is as follows:
\[J =\left|
\begin {array}{ccc}
    0&\g&\b
\\ \g&0&\a
\\ \b&\a&0
\end {array}
\right| = \left|\begin {array}{ccc}
 \ bc-\alpha^2&\a\beta-c\g&\alpha\gamma-b\beta
\\ \alpha\beta-c\gamma&ac-\beta^2&\beta\gamma-a\alpha
\\ \alpha\gamma-b\beta&\beta\gamma-a\alpha&ab-\gamma^2
\end {array}
\right|.\]
The condition $J\cross J=0$ gives the following relations amongst $\rho$, $\sigma$ and $\tau$:
\[\rho^2=\sigma^2=\tau^2=0, \hspace{10pt} \rho\sigma=\rho\tau=\sigma\tau=0.\]
We list the equations to be solved as follows:
\[ 2(\alpha\beta\gamma)=a\alpha^2+b\beta^2+c\gamma^2-abc, \]\[ \alpha^2=bc, \hspace{10pt}\beta^2=ac, \hspace{10pt}\gamma^2=ab, \hspace{10pt} c\ne 0\]
\[\tau=\alpha\beta-c\gamma ,\hspace{10pt} \rho = \beta\gamma - a\alpha, \hspace{10pt}\sigma = \alpha\gamma - b\beta.\]
If also $\tau = 0$, then $\gamma = \frac{1}{c}\alpha\beta$, so $\rho = \beta(\frac{1}{c}\alpha\beta) - a\alpha = \frac{1}{c}\alpha(\beta^2 - ac) = 0$ and 
$\sigma = \alpha(\frac{1}{c}\alpha\beta) - b\beta = \frac{1}{c}\beta(\alpha^2 - bc) = 0$, so $J = 0$, a contradiction.  So there are no solutions of this type
if $\tau = 0$.  So we take $\tau \ne 0$.  Then we have:
\[ \beta\tau = \beta(\alpha\beta-c\gamma) = \alpha\beta^2 - c\beta\gamma = \alpha ac - c(\rho + a\alpha) = - c\rho.\]
Since $\tau \ne 0 $, the general solution of the equation $\beta \tau = -c\rho$ has the form:
\[ \beta=-\rho\tilde{\tau}-\tau\tilde{\rho}\]
where $2\tau.\tilde{\tau} = c$ and $\tilde{\rho}$ is arbitrary.  Then the equation $\beta^2 = ac$ gives the formula: $2\rho.\tilde{\rho} = a$.
Similarly we find:
\[ \alpha\tau = \alpha^2\beta - c\alpha\gamma = c(\beta - \alpha\gamma) = - c\sigma, \]
\[ \alpha = - \tau\tilde{\sigma} - \sigma\tau', \hspace{10pt}2\tau.\tau' = c,\hspace{10pt} 2\sigma.\tilde{\sigma} = b.\]
Finally $\gamma$ is given in terms of $\alpha$ and $\beta$ by the formula:
\[ \gamma = \frac{1}{c}(\alpha\beta - \tau).\]
Then it may be checked that all the required equations are satisfied. We count the degrees of freedom in this solution as follows.  A priori there are $4n$ for
the quantities
$\tilde{\rho}$, $\tilde{\sigma}$, $\tilde{\tau}$ and
$\tau'$, but their variations (represented by the operator $\delta$), giving the same solution, are subject to the variety of constraints:
\[ \tau.\delta\tilde{\tau} = 0, \hspace{10pt}\tau.\delta\tau' = 0, \hspace{10pt} \rho.\delta\tilde{\rho} = 0,
\hspace{10pt}\sigma.\delta\tilde{\sigma} = 0,\]\[\rho\delta\tilde{\tau} + \tau\delta\tilde{\rho} = 0, \hspace{10pt}\tau \delta\tilde{\sigma}
+
\sigma\delta\tau' = 0.\]  Choose $\delta\tilde{\tau}$ and 
$\delta\tau'$ first, subject to only their two scalar constraints, giving $2n - 2$ degrees of freedom.  Then the solutions of the last two
equations for $\delta\tilde{\rho}$ and
$\delta\tilde{\sigma}$ automatically obey their scalar constraints, so they give an additional $2(\frac{n}{2}) = n$ degrees of freedom.  So the
degrees of freedom in the required solution amount to  $4n - ((2n - 2) + n) = n + 2$. 
\item The remaining case is the case that $a = b = c = 0$.  The conditions on the elements of $J$ are still $\rho^2=\sigma^2=\tau^2=0,
\hspace{10pt}\rho\sigma = 0,\hspace{10pt}\rho\tau =0,\hspace{10pt}\sigma\tau=0$.  The equation $K\times K = 2J$ now amounts to the following system:
\[ \alpha^2=\beta^2=\gamma^2=0, \]
\[ \tau=\alpha\beta,\hspace{10pt}\sigma=\alpha\gamma,\hspace{10pt}\rho=\beta\gamma, \hspace{10pt} (\alpha\beta\gamma) = 0.\]
Using the permutation symmetry, since $J\ne 0$, we may assume that $\tau \ne 0$, so each of $\alpha $ and $\beta$ must be non-zero also.  Then, since
$\alpha\tau = 0$, we have
$\alpha =
\tau\beta'$, where
$2\beta.\beta' = 1$.  Also, since $\rho = \beta\gamma$ and $\rho\beta = 0$, we have $\gamma - \rho \beta' = \beta\rho'$, for some $\rho'$, such that
$\rho'.\rho = 0$ (in order that
$\gamma^2 = 0$).  Then we have further:
\[ \sigma = \alpha\gamma =  - \rho'\tau + 2(\beta'\rho'\tau)\beta, \hspace{10pt}\sigma.\beta' = 0.\]
For $t$ an arbitrary real number, the equation $\rho'\tau = -\sigma + 2t\beta$ has the general solution $\rho' = -\sigma\tau' + 2t\beta\tau' +
\tilde{\beta}\tau$, where
$2\tau.\tau' = 1$ and $\tilde{\beta}$ is arbitrary.  Then we have $(\beta'\rho'\tau) = t$, so $\sigma =  - \rho'\tau + 2(\beta'\rho'\tau)\beta$, as
required.  Finally we need
$\beta$ to obey the equations $\beta\tau = 0$ and $\beta\rho = 0$.  So $\beta = \tau\alpha'$, where $2\alpha.\alpha' = 1$, $\rho.\alpha' = 0$ and
$2(\tau\alpha'\beta') =1$.  \eject\noindent Then we have: 
\[ \rho' = -\sigma\tau' + 2t\beta\tau' +
\tilde{\beta}\tau = -\sigma\tau' + 2t(\tau\alpha')\tau' +
\tilde{\beta}\tau \]\[ = -\sigma\tau' + 2t\alpha' + \beta''\tau, \hspace{10pt}\beta'' = \tilde{\beta} - 2t\tau'\alpha', \]
\[ \gamma = \rho \beta' + \beta\rho' = \rho \beta'  - \beta(\sigma\tau') + 2t\beta\alpha' + \beta(\beta''\tau)\]
\[ = \rho \beta' + \sigma(\beta\tau') + 2t(\alpha')^2\tau + 2\beta.\beta''\tau\]
\[ = \rho \beta' + \sigma((\tau\alpha')\tau') + 2t(\alpha')^2\tau + 2\beta.\beta''\tau\]
\[ =  \rho \beta' - \sigma((\tau'\alpha')\tau) + \alpha'\sigma + 2t(\alpha')^2\tau + 2\beta.\beta''\tau\]
\[ =  \rho \beta' + \alpha'\sigma + u\tau, \]
\[ u = 2(t(\alpha')^2- (\sigma\tau'\alpha') + \beta.\beta'').\]
Summarizing, we have, for some vectors $\alpha'$, $\beta'$ and scalar $u$, the general solution:
\[ \alpha = \tau\beta', \hspace{10pt} \beta = \tau\alpha', \hspace{10pt} \gamma =\rho \beta' + \alpha'\sigma + u\tau.\]
Back substituting, we find that these formulas give solutions of the required equations iff $\sigma.\beta' = \rho.\alpha' = 0$ and
$2(\tau\alpha'\beta') = 1$.  In particular $u$ is an arbitrary real number. When $\sigma$ and $\rho$ are non-zero,  there are $2n - 2$ degrees of
freedom in the choice of the quantities
$\alpha'$, $\beta'$ and $u$, but changing $\beta'$ and $\alpha'$ by multiples of $\tau$ with an appropriate change in $u$, gives the same solution for
$(\alpha, \beta, \gamma)$, so the number of degrees of freedom in the solution $(\alpha,\beta,\gamma)$ is $2n - 2 - 2(\frac{n}{2}) =  n - 2$.  If
exactly one of $\rho$ and $\sigma
$ vanish, then we add one degree of freedom, for  a total of $n -1$ and if both $\rho = \sigma = 0$, then we add two degrees of
freedom, giving a total of $n$.\end{itemize}\eject\noindent
\section{The Freudenthal symplectic form}
The set of pairs $(J,K)$ with $J$ in the Jordan algebra and $K$ in its dual has a natural symplectic two-form: $\omega_0 = dJ.dK$.  If $H$ is
any function of the pair $(J, K)$, the associated Hamiltonian vector field is $H' = H_J.\partial_K - H_K.\partial_J$, where the subscripts for $H$ denote
partial derivatives.  If $M$ is another such function, then we have the Poisson bracket: $\{H,M\} = H'(M) = H_J.M_K - H_K.M_J = - \{M,H\}$.  
Naturally associated to the symplectic form are the Lagrangian embeddings:
$2sJ = K\times K$ and
$2tK = J\times J$.  Here
$s$ and
$t$ are non-zero constants.  These are easily seen to be Lagrangian by evaluating the two-form $\omega_0$ directly.  Over the open set $\det(K) \ne 0$,
these two embeddings coincide provided that we have $\det(K) = s^2t$ and $\det(J) = st^2$.  We may then introduce the following naturally
defined scalar Hamiltonians, with their associated Hamiltonian vector fields:
\[ H_0 = J.K,\hspace{10pt}H_0' = K.\partial_K - J.\partial_J, \]
\[ H_1 = \det(J), \hspace{10pt}H_1' = \frac{1}{2}(J\times J).\partial_K,\]
\[ H_2 = \det(K),  \hspace{10pt} H_2' =  - \frac{1}{2}(K\times K).\partial_J,\]
\[ H_3 = (J\times J).(K\times
K), \hspace{10pt} H_3' =  2(K\times K).(J\times \partial_K) -   2(J\times J).(K\times \partial_J).\]
We obtain the following formulas for their Poisson brackets, showing that the Hamiltonians form a closed algebra under the Poisson bracket:
\[ \{H_0, H_1\} = -3H_1, \hspace{10pt}  \{H_0, H_2\} = 3H_2, \hspace{10pt} \{H_0, H_3\} = 0, \hspace{10pt} \{H_1, H_2\} = \frac{1}{4} H_3,\]
\[ \{H_1, H_3\} =  ((J\times J)\times (J\times J)).K = 8\det(J)J.K = 8H_0H_1, \]\[ \{H_2, H_3\} =   -((K\times K)\times (K\times K)).J =
-8\det(K)J.K = - 8H_0H_2.\]
We may also introduce the following quadratic Hamiltonians, depending on arbitrary constants $A$ and $B$:
\[ J_A = A.(J\times J), \hspace{10pt}  J'_A = 2(A\times J).\partial_K,\]
\[ K_B = B.(K\times K), \hspace{10pt}  K'_B = -2(B\times K).\partial_J, \]
\[ E_{A,B} = (A\times J).(B\times K), \hspace{10pt}  E'_{A,B} = (B\times K).(A\times\partial_K) -  (A\times J).(B\times\partial_J) .\]
Then we have the following Poisson brackets, valid for arbitrary constant $A$, $B$, $C$ and $D$:
\[ \{H_0, J_A\} = - 2J_A, \hspace{10pt}\{H_0, K_B\} = 2K_B, \hspace{10pt}\{H_0, E_{A, B}\} = 0, \]
\[ \{J_A, J_C\} = 0, \hspace{10pt} \{K_B,K_D\} = 0, \]
\[ \{J_A, K_B\} = 4E_{A,B},\]
\[ \{J_A, E_{C,D}\} = 2((A\times J)\times(C\times J)).D\]\[ = -  ((A\times C)\times(J\times J)).D + 2(A, C, J)J.D + (A, J, J)C.D + (C, J, J)A.D,\]
\[ =  -  J_{((A\times C)\times D - C.D A - A.D C)} + 2(A,C, J)(J.D).\]
At this point we see that the quadratic algebra does not close nicely (although, of course, the algebra of \emph{all} quadratic Hamiltonians does
close).  The present algebra is somewhat rigid, in that the Lagrangian embeddings are not scale invariant: the scales of $J$ and $K$ are tied
together.  The remedy is to pass to a larger space, invented by Freudenthal, with two extra real dimensions, denoted
$\alpha
$ and
$\beta$, making $56$ real dimensions in all, equipped with the symplectic form:
\[ \omega = dJ.dK + d\alpha d\beta.\]
Associated to any smooth functions $H$ and $M$ of the variables $(J, K, \alpha, \beta)$ are the Hamiltonian vector field and Poisson bracket:
\[ H' = H_J.\partial_K - H_K.\partial_J + H_\alpha\partial_\beta - H_\beta\partial_\alpha, \]
\[ \{H,M\} = H'M =  H_J.M_K - H_K.M_J + H_\alpha M_\beta - H_\beta M_\alpha.\]
Again we have natural Lagrangian manifolds, given by the following compatible system of equations:
\[ 2\alpha J = K\times K, \hspace{10pt} 2\beta K = J\times J, \hspace{10pt} \det(J) = \alpha\beta^2, \hspace{10pt}\det(K) = \beta\alpha^2.\]
Generalizing our previous Hamiltonians, we introduce the following four scalar Hamiltonians, together with their associated
Hamiltonian vector fields:
\[ H_0 = J.K - 3\alpha\beta,\hspace{10pt}H_0' = K.\partial_K - J.\partial_J - 3(\beta\partial_\beta - \alpha\partial_\alpha), \]
\[ H_1 = \det(J) - \alpha\beta^2, \hspace{10pt}H_1' = \frac{1}{2}(J\times J).\partial_K - \beta(\beta\partial_\beta - 2\alpha\partial_\alpha),\]
\[ H_2 = \det(K) - \beta\alpha^2,  \hspace{10pt} H_2' =  - \frac{1}{2}(K\times K).\partial_J - \alpha(2\beta\partial_\beta -
\alpha\partial_\alpha),\]
\[ H_3 = (J\times J).(K\times
K) - 12\alpha^2\beta^2, \]\[ H_3' =  2(K\times K).(J\times \partial_K) -   2(J\times J).(K\times \partial_J) -
24\alpha\beta(\beta\partial_\beta - \alpha\partial_\alpha).\]
The Freudenthal quartic invariant, $F$, in this language is the function on the phase space: $F = (J.K - \alpha\beta)^2 - (J\times J).(K\times K)  +
4\alpha\det(J) + 4\beta\det(K)$.  It may be expressed in terms of our Hamiltonians by the formula:
\[ F = - H_3 + H_0^2 + 2\alpha\beta H_0 + 4\alpha H_1 + 4\beta H_2.\]
In particular the Freudenthal invariant vanishes whenever the Hamiltonians $H_0$, $H_1$, $H_2$ and $H_3$ simultaneously vanish.  The Poisson
brackets are now as follows:
\[ \{H_0, H_1\} = -3H_1, \hspace{10pt}  \{H_0, H_2\} = 3H_2, \hspace{10pt} \{H_0, H_3\} = 0, \hspace{10pt} \{H_1, H_2\} = \frac{1}{4} H_3,\]
\[ \{H_1, H_3\} =  8\det(J)J.K   -24 \alpha^2\beta^3 \]
\[ =   8H_0H_1 + 8\alpha\beta^2 H_0 + 24\alpha\beta H_1,\]
\[ \{H_2, H_3\} =    - 8\det(K)J.K   + 24 \alpha^2\beta^3\]
\[ =  - 8H_0H_2 - 8\alpha^2\beta H_0 - 24\alpha\beta H_2,\]
As before, we may introduce the following quadratic Hamiltonians, depending on arbitrary constants $A$ and $B$:
\[ J_A = A.(J\times J - 2\beta K), \hspace{10pt}  J'_A = 2(A\times J).\partial_K + 2A.K\partial_\alpha + 2\beta A.\partial_J,\]
\[ K_B = B.(K\times K - 2\alpha J), \hspace{10pt}  K'_B = -2(B\times K).\partial_J - 2B.J\partial_\beta - 2\alpha B.\partial_K, \]
\[ E_{A,B} = (A\times J).(B\times K) - A.KB.J - \alpha\beta A.B, \]\[ E'_{A,B}\hspace{-1.5pt} = \hspace{-1.5pt} (B\hspace{-1.5pt} \times \hspace{-1.5pt}
K).(A\hspace{-1.5pt} \times\hspace{-1.5pt} \partial_K)\hspace{-1.5pt}  - \hspace{-1.5pt} (A\hspace{-1.5pt} \times\hspace{-1.5pt}  J).(B\hspace{-1.5pt}
\times\hspace{-1.5pt} \partial_J)\hspace{-1.5pt}  +\hspace{-1.5pt}  B.JA.\partial_J \hspace{-1.5pt} -\hspace{-1.5pt}  A.KB.\partial_K\hspace{-1.5pt} 
-\hspace{-1.5pt}  A.B(\beta\partial_\beta\hspace{-1.5pt}  -\hspace{-1.5pt} 
\alpha\partial_\alpha) .\] Notice that even when $\alpha$ and $\beta$ are both zero, the Hamiltonian
$E_{A,B}$ does not coincide with our earlier version.  Then by direct computation, using the various Jordan algebra identities proved above, we have the
following Poisson brackets, valid for arbitrary constant
$A$,
$B$,
$C$ and
$D$:
\[ \{J_A, J_C\} = 0, \hspace{10pt} \{K_B,K_D\} = 0, \]
\[ \{J_A, K_B\} = 4E_{A,B},\]
\[ \{J_A, E_{C,D}\}  =  -  J_{((A\times C)\times D - D.C A - D.A C)} \]
\[ \{K_B, E_{C,D}\}  =  K_{((B\times D)\times C - C.D B - C.B D)} \]
\[ \{E_{A,B}, E_{C,D}\} = E_{(A, \hspace{7pt}(B\times D)\times C - C.D B - C.B D)}- E_{((A\times C)\times D - D.C A - D.A C, \hspace{7pt}B)}. \]  
Remarkably for the Freudenthal symplectic form, this algebra closes, giving a Lie algebra, in sharp contrast to the corresponding algebra for the
symplectic form
$dJ.dK$.   For the case of either the octaves or  the split octaves, the Lie algebra has dimension $133$ and is a non-compact real form of the
complex Lie algebra
$\E_7$.  The subalgebra generated by the
$E_{A,B}$ Hamiltonians has dimension $79$ and is the direct sum of the one dimensional grading algebra generated by the Hamiltonian $H_0$ and a 78
dimensional algebra, which is a non-compact real form of the complex Lie algebra $\E_6$.  For the case of the complex octaves, the corresponding
algebras are the complex Lie algebras $\E_7$ and (the sum of a one-dimensional complex algebra and) $\E_6$ themselves.  Finally note that on the
Lagrangian submanifold, all the Hamitonians vanish, including the Freudenthal quartic invariant.  Also note that the calculations here apply to any
Jordan algebra of the three by three matrix type: so for example in the case that we use the algebra of real symmetric three by three matrices, the Lie
algebra we obtain is the algebra
$\OO(3,4)$.
\eject\noindent
\section{The Jordan algebra in the $\mathcal{O}(5,5)$ formalism}
Using $\mathcal{O}(5,5)$ spinors and vectors, the Jordan matrix may be represented by a scalar, a vector and a chiral spinor:
\[ J = \hspace{5pt}\begin{array}{|ccc|}a&\gamma&\beta\\\gamma&b&\alpha\\\beta&\alpha&c\end{array}  \hspace{3pt}\rightarrow \hspace{3pt}(b,
\underline{X}, \psi ) \hspace{3pt}= \hspace{3pt}(b, (c,a,\beta),
\begin{array}{|c|}\alpha\\\gamma\end{array}\hspace{3pt}).\]
Then we have $\psi'\underline{\gamma}\psi = (\alpha^2, \gamma^2, \alpha\gamma)$ and $2\psi'X\psi = 2\underline{X}.\psi'\underline{\gamma}\psi =
c\gamma^2 + a\alpha^2 - 2(\alpha\beta\gamma)$. Also $\underline{X}^2 = ac - \beta^2$. So, for the Jordan determinant we have:
\[ \det(J) = b\underline{X}^2-2\psi'X\psi.\]
The information in the Jordan cross product $J\rightarrow J\times J$  is identical to the information in the following map:
\[ J = (b, \underline{X}, \psi) \rightarrow (-\underline{X}^2, \psi'\underline{\gamma}\psi - b\underline{X}, X\psi).\]
We may check this by writing out the map explicitly:
\[  (b, (c,a,\beta), \psi) \rightarrow (\beta^2 - ac, (\alpha^2 - bc, \gamma^2 - ba, \alpha\gamma -
b\beta),\frac{1}{2}\hspace{3pt}\begin{array}{|c|}a\alpha -
\beta\gamma\\c\gamma - \alpha\beta\end{array}\hspace{3pt}).\]
We see that all the terms of the Jordan product occur, albeit with slightly different coefficients.  Also note that in this description, the image lies
in the dual space, since $X\psi$ has the opposite chirality to that of $\psi$.  We may work with both chiralitites at once, by simply dropping the
condition that
$\psi$ be chiral and retaining the same formula for the map, which also makes sense in any dimension. Then if we iterate the map we find the formula:
\[ J = (b, \underline{X}, \psi) \rightarrow (-\underline{X}^2, \psi'\underline{\gamma}\psi - b\underline{X}, X\psi) \rightarrow (c,
\underline{Y}, \phi), \]
\[ c = -( \psi'\underline{\gamma}\psi - b\underline{X})^2 = - ( \psi'\underline{\gamma}\psi)^2 - b( b\underline{X}^2-2\psi'X\psi) \]\[= - (
\psi'\underline{\gamma}\psi)^2 - b\det(J),\]
\[ \underline{Y} =  \psi'X\underline{\gamma}X\psi +\underline{X}^2(\psi'\underline{\gamma}\psi - b\underline{X}) = - \det(J)\underline{X}, \]
\[ \phi = (\psi'\underline{\gamma}\psi - b\underline{X}).\underline{\gamma} X\psi =  (\psi'\underline{\gamma}\psi.(2\underline{\X} -
X\underline{\gamma}\psi) - b(\underline{X})^2\psi = -
X(\psi'\underline{\gamma}\psi).(\underline{\gamma}\psi) - \det(J)\psi.\]
\eject\noindent  So we have after two iterations:
\[ J\rightarrow -\det(J)J - ((\psi'\underline{\gamma}\psi)^2, 0, X(\psi'\underline{\gamma}\psi).(\underline{\gamma}\psi) ).\]
Next, computing $\det(J\times J)$, we get:
\[ \det(J\times J) = -X^2(\psi'\underline{\gamma}\psi - b\underline{X})^2 - 2\psi'X(\psi'\underline{\gamma}\psi -
b\underline{X}).\underline{\gamma}X\psi\]
\[ =  -X^2(\psi'\underline{\gamma}\psi)^2 - b^2(X^2)^2 + 4bX^2\psi'X\psi - 2\psi'X(\psi'\underline{\gamma}\psi).(2\underline{X}
- X\underline{\gamma})\psi\]
\[ = -(\det(J))^2 + X^2(\psi'\underline{\gamma}\psi)^2.\]
Summarizing, we have:
\[ J = (b, \underline{X}, \psi), \hspace{10pt}\det(J) = bX^2-2\psi'X\psi,\]
\[ J\times J = (-X^2, \psi'\underline{\gamma}\psi - b\underline{X}, X\psi),\]
\[ \det(J\times J) = -(\det(J))^2 + X^2(\psi'\underline{\gamma}\psi)^2, \]
\[ (J\times J)\times(J\times J) =  -\det(J)J - ((\psi'\underline{\gamma}\psi)^2, 0, X(\psi'\underline{\gamma}\psi).(\underline{\gamma}\psi) ).\]
In ten dimensions, we say that $J$ is chiral iff $\psi $ is chiral.  Specializing to the \emph{chiral} case, in ten dimensions, the vector
$\psi\underline{\gamma}\psi$ is null and orthogonal to
$\underline{\gamma}\psi$.  Then the basic formulas become:
\[ J = (b, \underline{X}, \psi), \hspace{10pt}\det(J) = bX^2-2\psi'X\psi,\]
\[ J\times J = (-X^2, \psi'\underline{\gamma}\psi - b\underline{X}, X\psi),\]
\[ \det(J\times J) = -(\det(J))^2, \]
\[ (J\times J)\times(J\times J) =  -\det(J)J.\]
Next, polarizing the formula for
$J\times J$ and putting
$K = (c,
\underline{Y}, \phi)$, we have:
\[ 2J\times K = (-2X.Y, 2\psi'\underline{\gamma}\phi - b\underline{Y} - c\underline{X}, X\phi + Y\psi).\]
In particular, we have:
\[ 2J\times (J\times J) = (-2\psi'X\psi + 2bX^2,(b^2 +  2\psi'\psi + X^2)\underline{X} - b\psi'\underline{\gamma}\psi,
X^2\psi + (\psi'\underline{\gamma}\psi).\underline{\gamma}\psi - bX\psi)\]
\[ = - b(J\times J)  + X^2J + (-2\psi'X\psi, 2\psi'\psi\underline{X}, (\psi'\underline{\gamma}\psi).\underline{\gamma}\psi).\]
Now put also $L = (d, \underline{Z}, \omega)$.  Then polarizing the determinant gives a trilinear form $(J, K, L)$, such that $(J,J,J) = 3\det(J)$:
\[ (J,K,L) = bY.Z + cZ.X + dX.Y - 2\psi'Y\omega - 2 \psi'Z\phi - 2\phi'X\omega, \hspace{10pt} (J,J,J) = 3\det(J).\]
We may write the triple product as $J.(K\times L)$, where, if $J_1 = (b_1, \underline{X}_1, \psi_1)$ and  $J_2 = (b_2, \underline{X}_2, \psi_2)$, we
have the dot product:  $J_1.J_2 = - b_1b_2 - 2X_1.X_2 - 4\psi_1'\psi_2 $.\\
Then we have:
\[ J.J = -b^2 - 2X^2 - 4\psi'\psi, \]
\[ J.(J\times J) = 3bX^2 - 6\psi'X\psi = 3\det(J), \]
\[ (J\times J).(J\times J) = - X^4- 2(\psi'\underline{\gamma}\psi - b\underline{X})^2 - 4(\psi'XX\psi)\]
\[ = - X^4 - 4X^2\psi'\psi - 2(\psi'\underline{\gamma}\psi)^2 - 2b\det(J).\]
In the chiral case, we have:
\[ \frac{1}{2}( - (J.(J\times J))^2 + J.J(J\times J).(J\times J)) =\frac{1}{2}( (X^4 + 2b\det(J))(b^2 + 2X^2) - 9\det(J)^2)\]
\[ = X^6 + 2bX^2\det(J))  +  b^2/2X^4 + b^3\det(J)- \frac{9}{2}\det(J)^2.\]
\eject\noindent
\section{The split octaves and twistors}
We consider henceforth the case that $\V$ is twenty-four dimensional, so that each $\V_i$ is eight dimensional and we require that the metric of each $\V_i$
be of neutral signature, so that each $\V_i$ has the symmetry group $\mathcal{O}(4,4)$.  The group $\mathcal{O}(4,4)$ contains the group $\mathcal{U}(2,2)$,
which acts naturally by complex linear transformations on a four complex dimensional vector space, called twistor space, $\T$, preserving a pseudo-hermitian
form of signature $(2,2)$.  Using abstract indices, a twistor $Z$ may be represented as $Z^\alpha$, its conjugate by $\overline{Z}_\alpha$ and its
(indefinite) norm squared by $Z^\alpha \overline{Z}_\alpha$.  This conjugation extends naturally to the tensor algebras associated to $\T$.  Fix an
alternating volume form for $\T$, $\epsilon_{\alpha\beta\gamma\delta}$, normalized against its conjugate $\epsilon^{\alpha\beta\gamma\delta}$ by the
formula $\epsilon_{\alpha\beta\gamma\delta}\epsilon^{\alpha\beta\gamma\delta} = 24$.\\\\Our three vector spaces are now as follows:
\begin{itemize}\item $\V_0 = \T_0 = \{(x, X^{\alpha\beta}): x\in \C, X^{\alpha\beta} = - X^{\beta\alpha} =
\frac{1}{2}\epsilon^{\alpha\beta\gamma\delta}\overline{X}_{\gamma\delta} \in \Omega^2(\T)\}$.  \\We give the space $\V_0$ the norm squared function
$x\overline{x} + \frac{1}{4}X^{\alpha\beta}\overline{X}_{\alpha\beta}$. It is easily verified that this norm has signature $(4,4)$.  Note that $\V_0$ is
naturally a real vector space, but has no natural action of the complex numbers on it, even though $\V_0$ is defined using complex numbers. 
\item $\V_1 = \T$, equipped with its standard pseudo-hermitian form.
\item $\V_{-1} = \T$,  equipped with its standard pseudo-hermitian form.
\end{itemize} 
To distinguish the two twistor spaces involved, we will denote them by $\T_{\pm}$, with $\T_{+}$ for $\V_1$ and $\T_{-}$ for $\V_{-1}$.  The three real
bilinear multiplications are now as follows:
\begin{itemize}\item Given $Z^\alpha \in \T_{+}$ and $(x, X^{\alpha\beta}) \in \T_0$, their product $(Z, (x,X)) = ((x, X), Z)$ in $\T_{-}$ is:
\[ (Z, (x, X))^\alpha = xZ^\alpha + X^{\alpha\beta}\overline{Z}_\beta.\]
\item Given $Y^\alpha \in \T_{-}$ and $(x, X^{\alpha\beta}) \in \V_0$, their product $(Y, (x, X)) = ((x, X), Y)$ in $\T_{+}$ is:
\[ (Y, (x, X))^\alpha = \overline{x}Y^\alpha - X^{\alpha\beta}\overline{Y}_\beta.\]
\item Given $Z^\alpha \in \T_{+}$ and $Y^\alpha \in \T_{-}$, their product $(Z, Y) = (Y, Z)$ in $\V_{+}$ is:
\[ (Z, Y) = (Y^\gamma \overline{Z}_{\gamma}, 2Y^{[\alpha}Z^{\beta]} + \epsilon^{\alpha\beta\gamma\delta}\overline{Y}_\gamma\overline{Z}_\delta).\]
Here, and in the following, we use square brackets around tensor indices to indicate idempotent skew-symmetrization: so for example $ 2Y^{[\alpha}Z^{\beta]}
= Y^\alpha Z^\beta - Z^\alpha Y^\beta$.
\end{itemize}
We verify by direct calculation that these products are norm preserving:
\begin{itemize}\item First the quantity $(Z, (Z, Y))$:\[(Z, (Z, Y)) = (Z, (Y^\alpha Z_{\alpha}, 2Y^{[\alpha}Z^{\beta]} +
\epsilon^{\alpha\beta\gamma\delta}\overline{Y}_\gamma\overline{Z}_\delta))) \]\[= Y^\gamma \overline{Z}_{\gamma}Z^\alpha + (2Y^{[\alpha}Z^{\beta]} +
\epsilon^{\alpha\beta\gamma\delta}\overline{Y}_\gamma\overline{Z}_\delta )\overline{Z}_\beta\]
\[ =  Y^\gamma \overline{Z}_{\gamma}Z^\alpha + Y^{\alpha}Z^{\beta}\overline{Z}_\beta - Z^{\alpha} Y^{\beta}\overline{Z}_\beta
= (Z^{\beta}\overline{Z}_\beta) Y^{\alpha}. \] 
\item Next the quantity $(Z, (Z, (x,X)))$:
\[ (Z, (Z, (x,X))) = (Z^\gamma, xZ^\alpha + X^{\alpha\beta}\overline{Z}_\beta)\]
\[ = ((xZ^\alpha + X^{\alpha\beta}\overline{Z}_\beta)\overline{Z}_{\alpha},  2(xZ^{[\alpha} - X^{\gamma[\alpha}\overline{Z}_\gamma) Z^{\beta]} +
\epsilon^{\alpha\beta\gamma\delta}(\overline{x}\overline{Z}_{\gamma} + \overline{X}_{\gamma\sigma}Z^\sigma )\overline{Z}_\delta)\]
\[ = (xZ^\alpha\overline{Z}_{\alpha}, - 2\overline{Z}_\gamma X^{\gamma[\alpha} Z^{\beta]} + \frac{1}{2}
\epsilon^{\gamma\alpha\beta\delta}\epsilon_{\gamma\sigma\rho \tau}X^{\rho\tau}Z^\sigma \overline{Z}_\delta)\]
\[ = (xZ^\alpha\overline{Z}_{\alpha}, - 2\overline{Z}_\delta X^{\delta[\alpha} Z^{\beta]} + 3X^{[\alpha\beta}Z^{\delta]} \overline{Z}_\delta)\]
\[ = (Z^\alpha\overline{Z}_{\alpha})(x, X).\]
\item Next the quantity $(Y, (Y, Z))$:\[(Y, (Y, Z)) = (Y, (Y^\alpha Z_{\alpha}, 2Y^{[\alpha}Z^{\beta]} +
\epsilon^{\alpha\beta\gamma\delta}\overline{Y}_\gamma\overline{Z}_\delta))) \]\[= Z^\gamma \overline{Y}_{\gamma}Y^\alpha - (2Y^{[\alpha}Z^{\beta]} +
\epsilon^{\alpha\beta\gamma\delta}\overline{Y}_\gamma\overline{Z}_\delta )\overline{Y}_\beta\]
\[ =   Z^\gamma \overline{Y}_{\gamma}Y^\alpha - Y^{\alpha}Z^{\beta}\overline{Y}_\beta + Z^{\alpha} Y^{\beta}\overline{Y}_\beta
= (Y^{\beta}\overline{Y}_\beta) Z^{\alpha}. \] 
\item Next the quantity $(Y, (Y, (x,X)))$:
\[ (Y, (Y, (x,X))) = (Y^\gamma, \overline{x}Y^\alpha - X^{\alpha\beta}\overline{Y}_\beta)\]
\[ = (Y^\alpha(x\overline{Y}_\alpha - \overline{X}_{\alpha\beta}Y^\beta),  2Y^{[\alpha}(\overline{x}Y^{\beta]} - X^{\beta]\gamma}\overline{Y}_\gamma) +
\epsilon^{\alpha\beta\gamma\delta}\overline{Y}_{\gamma}( x\overline{Y}_\delta - \overline{X}_{\delta\sigma}Y^\sigma))\]
\[ = (xY^\alpha\overline{Y}_{\alpha}, -  2Y^{[\alpha} X^{\beta]\gamma}\overline{Y}_\gamma+ \frac{1}{2}
\overline{Y}_{\gamma}\epsilon^{\delta\alpha\beta\gamma}\epsilon_{\delta\sigma\rho
\tau}X^{\rho\tau}Y^\sigma)\]
\[ = (xY^\alpha\overline{Y}_{\alpha}, -  2Y^{[\alpha} X^{\beta]\gamma}\overline{Y}_\gamma + 3X^{[\alpha\beta}Y^{\gamma]} \overline{Y}_\gamma)\]
\[ = (Y^\alpha\overline{Y}_{\alpha})(x, X).\]
\item Next the quantity $((x,X), ((x,X), Z))$:
\[ ((x, X), ((x,X), Z)) = ((x, X), xZ^\alpha + X^{\alpha\beta}\overline{Z}_\beta)\]
\[ = \overline{x}(xZ^\alpha + X^{\alpha\beta}\overline{Z}_\beta) - X^{\alpha\gamma}(\overline{x}\overline{Z}_\gamma +
\overline{X}_{\gamma\delta}Z^\delta)\]
\[ = x\overline{x}Z^\alpha + X^{\alpha\gamma}
\overline{X}_{\delta\gamma}Z^\delta\]
\[ = ( x\overline{x} + \frac{1}{4}X^{\beta\gamma}
\overline{X}_{\beta\gamma})Z^\alpha.\]
\item Finally the quantity $((x,X), ((x,X), Y))$:
\[ ((x, X), ((x,X), Y)) = ((x, X), \overline{x}Y^\alpha - X^{\alpha\beta}\overline{Y}_\beta)\]
\[ = x(\overline{x}Y^\alpha - X^{\alpha\beta}\overline{Y}_\beta) + X^{\alpha\gamma}(x\overline{Y}_\gamma -
\overline{X}_{\gamma\delta}Y^\delta)\]
\[ = x\overline{x}Y^\alpha + X^{\alpha\gamma}
\overline{X}_{\delta\gamma}Y^\delta\]
\[ = ( x\overline{x} + \frac{1}{4}X^{\beta\gamma}
\overline{X}_{\beta\gamma})Y^\alpha.\]
\end{itemize}
The corresponding triple product is as follows:
\[ ((x, X), Y, Z) =\frac{1}{2}( xZ^\alpha\overline{Y}_\alpha + \overline{x}Y^\alpha\overline{Z}_\alpha+ X^{\alpha\beta}\overline{Y}_\alpha\overline{Z}_\beta +
\overline{X}_{\alpha\beta}Y^\alpha Z^\beta).\]
Returning to the Jordan algebra, the elements $J$ of the algebra may be written $J = (a,b,c, Z, (x, X), Y)$, with $a$, $b$ and $c$ real numbers.  Then the
Jordan cross product is:
\[ J \times J = 2(bc - Z^\alpha\overline{Z}_\alpha,\hspace{10pt} ca - x\overline{x} - \frac{1}{4}X^{\alpha\beta}\overline{X}_{\alpha\beta},\hspace{10pt} ab -
Y^\alpha Y_{\alpha},\]\[
\overline{x}Y^{\alpha} -
X^{\alpha\beta}\overline{Y}_\beta,\hspace{10pt} (Y.\overline{Z} - bx,\hspace{10pt} 2Y^{[\alpha}Z^{\beta]} + \epsilon^{\alpha\beta\gamma\delta}Y_\gamma
Z_\delta - b X^{\alpha\beta}), \hspace{10pt}xZ^\alpha + X^{\alpha\beta}\overline{Z}_\beta).\]
Finally the Jordan triple product is:
\[ (J, J, J) = 6(abc -  aZ^\alpha\overline{Z}_\alpha - b(x\overline{x} + \frac{1}{4}X^{\alpha\beta}\overline{X}_{\alpha\beta}) - cY^\alpha\overline{Y}_\alpha
\]\[+ xZ^\alpha\overline{Y}_\alpha + \overline{x}Y^\alpha\overline{Z}_\alpha+ X^{\alpha\beta}\overline{Y}_\alpha\overline{Z}_\beta +
\overline{X}_{\alpha\beta}Y^\alpha Z^\beta).\]\eject\noindent
\section{The internal $\mathcal{SL}(2, \C)$ symmetry; \\Planck's vector, angular momentum and mass}
In the previous section, the split octave Jordan algebra was constructed using ordinary twistor spaces for two of the triple of vector spaces.  Inspection
of the formulas reveals a slightly subtle parallelism between the way the two twistor spaces enter the various expressions.  It emerges that the
two twistor spaces assemble naturally into a complex two-component spinor, with $\mathcal{SL}(2, \C)$ structure group.  This $\mathcal{SL}(2, \C)$ acts
\emph{independently} of the
$\mathcal{SU}(2, 2)$ group acting on the twistors, so it will be called an internal symmetry group, the  $\mathcal{SU}(2, 2)$ group being external.

Relative to a primed spinor basis $(o^{A'}, \iota^{A'})$, normalized so that $o_{A'}\iota^{A'} = 1$, where indices are raised and lowered using the
alternating spinor symplectic form $\epsilon_{A'B'}$ and its conjugate $\epsilon_{AB}$, introduce the combination of the twistors
$Z^\alpha$ and $Y^\alpha$:
\[ Z^{\alpha B'} = Z^\alpha \iota^{B'} - Y^\alpha o^{B'}.\]
Also define a \emph{real Lorentz four-vector} $x^a$ in the internal symmetry space by:
\[ x^a = - ao^Ao^{A'} - c\iota^{A}\iota^{A'} + x\iota^Ao^{A'}  + \overline{x}o^A\iota^{A'}.\]
The Lorentzian length squared of the four vector $x^a$ is $x^ax_a = 2(ac - x\overline{x})$.  Then the twenty-seven parts of a Jordan
algebra element are packaged as follows:
\[ J = (b,  x^a,  Z^{\alpha B'}, X^{\alpha\beta}).\]
The Jordan cross product is now:
\[ J\times J = 2(\frac{1}{2}x^ax_a - \frac{1}{4}X^{\alpha\beta}\overline{X}_{\alpha\beta}, \hspace{10pt} bx^a + Z^{\beta
A'}\overline{Z}_{\beta}^{A},\hspace{10pt}  X^{\alpha\beta}\overline{Z}_\beta^B + x^B_{B'}Z^{\alpha B'}, \]\[ bX^{\alpha\beta} -
Z^{\alpha C'}Z^{\beta}_{C'} -
\frac{1}{2}\epsilon^{\alpha\beta\gamma\delta}\overline{Z}_\gamma^A\overline{Z}_{\delta A}).\]
Note that the cross product takes values in the dual space of the Jordan algebra: it has an extra chirality, not visible before.  A general dual element $K$
may be represented as:
\[ K =  (c,  y^a,  Y^{\alpha B}, Y^{\alpha\beta}).\]
Then the dual pairing is:
\[ J.K = bc + x^ay_a - Y^{\alpha B}\overline{Z}_{\alpha B} +  Z^{\alpha B'}\overline{Y}_{\alpha B'} -\frac{1}{2}X^{\alpha\beta}Y_{\alpha\beta}.\]
The triple product is now:  
\[ (J, J, J) = J.(J\times J)\]\[ = 3(b(x^ax_a - \frac{1}{2}X^{\alpha\beta}\overline{X}_{\alpha\beta}) + 2x_{b}Z^{\alpha B'}\overline{Z}_\alpha^B +
X^{\alpha\beta}\overline{Z}_\alpha^C\overline{Z}_{\beta C} + \overline{X}_{\alpha\beta}Z^{\alpha C'}Z^{\beta}_{C'}).\] 
Consider the action of the group $\U(2,2)$ on the Jordan algebra.  Representing the Lie algebra by an anti-hermitian system of
vector fields, $E_\beta^\alpha$, we have:
\[ E_\beta^\alpha = Z^{\alpha A'}\partial_{\beta \A'} - \overline{Z}_{\beta}^A\overline{\partial}^{\alpha}_A + 2X^{\alpha\gamma}\partial_{\beta\gamma}
- 2\overline{X}_{\beta\gamma}\overline{\partial}^{\alpha\gamma}.\]
If we think of these operators as obtained from Hamiltonians, then their twistor part is the real part of the quantity
$Z^{\alpha A'}\overline{Y}_{\beta A'}$, so is defined at the level of the Freudenthal phase space, where the system is part of the ensemble of Hamiltonians given
in section nine above.  However, when we pass to the Lagrangian submanifold, thus eliminating the twistors $Y^{\alpha A}$,  we still have the vector fields
representing the symmetry action, but we have lost the symplectic structure (by definition of Lagrangian) and  also the Hamiltonians, since they all
vanish on the Lagrangian manifold.  We would like to recover some kind of Hamiltonian structure.  A possible clue comes from the twistor quantization of
particles, which has each twistor canonically conjugate to its conjugate dual twistor.  So here we expect that $Z^{\alpha A'}$ and $\overline{Z}_{\beta}^B$ should
be canonically conjugate.  This entails a new symplectic form of the form $idZ^{\alpha A'}d\overline{Z}_{\alpha}^A$.  However this is impossible, unless we dispose
of the free spinor indices ${A'}$ and $A$ on this form.  With the ingredients available, there seems to be only one reasonable way to do this, giving the two-form:
\[ \omega = x_{AA'}dZ^{\alpha A'}d\overline{Z}_{\alpha}^A.\]
Using this form, we are able to write the canonical commutation relations for $Z^{\alpha A'}$ as follows:
\[ [Z^{\alpha A'}, \overline{Z}_{\beta}^A] = i x^{AA'}\delta_\beta^\alpha, \hspace{10pt} [Z^{\alpha A'}, Z^{\beta B'}] = 0.\]
From this point of view, \emph{Planck's constant} is part of the four-vector $x^a$, which we call \emph{Planck's vector}: it is actually the Lorentz length of
this vector!  Thus also from this viewpoint, the symmetry breaking procedure must be such that $x^a$, originally dynamical, becomes \emph{fixed} at a constant
value.  In other words Planck's vector is an \emph{order parameter}.  This raises the question whether other phases might be allowed where Planck's vector is
spacelike or null.
\eject\noindent At this level the twistor part of the Hamiltonians for the conformal algebra is
$E^\alpha_\beta = Z^{\alpha A'}\overline{Z}_\beta^A x_{AA'}$.  This agrees with the standard formula for a two twistor particle, when a basis
for the internal spin space is chosen so that $E^\alpha_\beta$ is proportional to the sum of the outer products of two twistors: $Z^\alpha\overline{Z}_\beta
+ W^\alpha\overline{W}_\beta$ \cite{Wo2}. The question as to why one should use the plus sign in the middle of the latter formula has bothered twistor
theorists for a long time.  In this scenario, the answer is deep indeed!  \\\\
Finally we contemplate the further reduction to the Poincare group to describe particles of a fixed mass.  Actually, we can consider, more generally, the
reduction to any of the three ten-dimensional subgroups of $\SU(2,2)$, relevant for conformally flat cosmologies: de Sitter (Spin(1,4)), Anti-de Sitter
(Spin(2,3)) and Poincare (the semi-direct product of 
$\mathcal{S}\mathcal{L}(2, \C)$ with the translations).  Each is the invariance group of a
skew two-index tensor in twistor space, $I_{\alpha\beta}$, such that
$I_{\alpha\beta} =
\frac{1}{2}\epsilon_{\alpha\beta\gamma\delta}\overline{I}^{\gamma\delta}$.   The three groups then arise according to the sign of the invariant
$I_{\alpha\beta}\overline{I}^{\alpha\beta}$: positive for $\textrm{Spin}(1,4)$, negative for $\textrm{Spin}(2,3)$ and zero for the Poincare group \cite{Pe2}. 
Corresponding to these groups is the associated energy momentum tensor, representing the Lie algebra of the group, a symmetric two-index twistor
$A_{\alpha\beta}$ given by the formula
$A_{\alpha\beta} = 2I_{\gamma(\alpha} E^\gamma_{\beta)} $ and obeying the reality condition: $A_{\alpha\gamma}\overline{I}^{\beta\gamma} =
\overline{A}^{\beta\gamma}I_{\alpha\gamma}$ (so it has ten real independent components, in general, as required).  Here, in view of the interpretation of
the four-vector
$x^a$ just given, it is irresistible to identify $I_{\alpha\beta}$ with $\overline{X}_{\alpha\beta}$, giving the following proposal for the energy-momentum
tensor:
\[ A_{\alpha\beta} =    2\overline{X}_{\gamma(\alpha} Z^{\gamma A'}\overline{Z}^A_{\beta)}x_{AA'}.\]     
Note that this formula uses the complete projective information in an element $J$ of the Jordan algebra: only the parameter $b$ is missing, which can be scaled out
(or possibly fixed by a natural condition such as $\det(J) = 0$).  It is perhaps this formula which epitomizes the theme of this work: in this interpretation,
the projective Jordan algebra elements represent all possible massive (and massless) energy-momentum tensors, for all possible worlds with any value of the
cosmological constant and with any value of Planck's vector; here the twistors (taking up sixteen dimensions) represent the particle concept and the other ten
dimensions the geometry and quantum mechanics.  So the symmetry breaking procedure, in passing from the triality invariant Jordan algebra level to the level of
geometry and quantum mechanics, presumably first fixes the ten-dimensional
$\textrm{Spin}(5,5)$ vector
$(X^{\alpha\beta}, x^a)$, then splits it into its two constituents, at this point giving rise to the structure around us and allowing for the idea of massive
particles.  The question of which is the correct group is a matter for experiment!   
\eject\noindent
\section{The proposed symmetry breaking scenario}
We discuss a series of possible transitions, taking us from the level of a real form of $\E_8$ down to the group
$\mathcal{S}\mathcal{U}(2,2)$ and to the Poincare group.  We would expect that each transition is driven by a phase transformation in an
antecedent of the Zhang-Hu quantum fluid.  We should note that something like this scenario has occurred to others beforehand, but here the physics involved is
apparently different \cite{Sm1}.
\begin{itemize}\item \textbf{The $\E_8$ level: the Gunaydin-Koepsell-Nicolai theory}\\
We begin by envisaging a theory at the level of the non-compact real form of the complex Lie group $\E_8$, used particularly in the work of Gunaydin,
Koepsell and Nicolai.  At this level there would be no traditional spacetime geometry and no quantum mechanics of the usual kind.  The basic space of
the theory would be the 57-dimensional Heisenberg algebra formed by the elements of negative grade in the algebra of $\E_8$ as given by Gunaydin.  Note
that in their Heisenberg algebra, the operator that might normally correspond to Planck's constant in the Heisenberg approach to quantum mechanics, is
indeed an operator and has no distinguished eigen-value.  Indeed this operator is actually part of a natural $\mathcal{S}\mathcal{L}(2, \R)$ structure
and at this level we would expect that $\mathcal{S}\mathcal{L}(2, \R)$ structure to replace quantum mechanics.  
\item \textbf{The $\E_7$ level: the Freudenthal phase-space dominates}\\ At this level we would separate out the $57$-dimensional space into a $56+1$
decomposition.  Then the Freudenthal symplectic structure comes into play.  This is the level of the Hamiltonian theory of section nine.  The
$\E_7$ algebra arises naturally as shown in the that section.  The group $\E_7$ in turn is just the group of automorphisms of the Freudenthal
symplectic space, equipped with the triple product.   Here, however, the two Jordan algebra components of the Freudenthal phase space have not yet
separated out.
\item \textbf{The $\E_6$ level: the Jordan algebra emerges}\\To go to the next level, we have  a definite proposal, based on the work section
nine:  the theory focusses down to the Lagrangian submanifold of the Freudenthal space, losing half its degrees of freedom in the process.  It is
noteworthy in this respect that all the various Hamiltonians described there vanish on the Lagrangian submanifold. Included here is the Freudenthal
quartic form, which plays a big role in the $\E_8$ theory of Gunaydin, Koepsell and Nicolai; an attractive possibility is that at the higher levels,
the fluid is a gas, which condenses at this level.  
\item \textbf{The $\textrm{Spin}(5,5)$ level: triality breaks down}\\Until this point the idea of a spinor has not yet crystallized fully: in order to
do so, we must break the triality invariance of the Jordan algebra, and now we can allow for the emergence of both spinors and geometry.  Presumably
only at this stage do ideas like supersymmetry become relevant, since they seem to require a spinorial formulation.  When $\E_6$ is broken down to
$\textrm{Spin}(5,5)$, the twenty-seven dimensions of the  Jordan algebra are broken into a scalar, a ten-dimensional vector and a sixteen dimensional
chiral spinor for $\textrm{Spin}(5,5)$.
\item \textbf{The $\textrm{Spin}(4,4)$ level: the emergence of twistor theory}\\At the level of $\textrm{Spin}(4,4)$, individual twistors exist, but
they have no natural complex structure.  Here the edge in twistor space, used by the Zhang-Hu theory can be defined for the first time.
\item \textbf{The $\textrm{SU}(2,2)$ level}\\ Now the twistors acquire their traditional complex structure and traditional four dimensional spacetime
can exist.
\item \textbf{The Poincare level} \\The last step is presumably the breaking of conformal invariance, allowing for traditional massive
particles to exist.  See section twelve above for more details on this step.
\end{itemize}
\section{Acknowledgements}We thank the Mathematics Department of the University of Pittsburgh for partially funding a conference where some of these
ideas were discussed.  We thank the group at Stanford University centered around Shou-Cheng Zhang for very valuable discussions,
particularly Shou-Cheng Zhang himself, Jiang-Ping Hu, Nicolaos Toumbas and Andrei Bernevig.  We also thank Lee Smolin.  We thank the other members of the
Laboratory of Axiomatics for their support, particularly Tom Metzger, Devendra Kapadia, David Hillman, Jonathan Holland, Suresh Maran, Dana Mihai, Erin
Sparling, Sarah Lewis, Saeeda Hafiz and Zed Armstrong.


\newpage\begin{thebibliography}{30}
\bibitem{Zh2}S.-C. Zhang and J.-P. Hu, \hspace{3pt}\emph{A Four-Dimensional Generalization of the Quantum Hall Effect},\hspace{3pt} Science
\textbf{294}(5543), 823-828, 2001. 
\bibitem{Zh6}B.A. Bernevig, C.-H. Chern, J.-P. Hu, N. Tuombas, and S.-C. Zhang,\hspace{3pt}  \emph{Effective Field Theory description of the higher dimensional
quantum Hall liquid},\hspace{3pt}  Annals of Physics, \textbf{300}, 185, 2002.
\bibitem{Ch2}Y.X. Chen,\hspace{3pt} \emph{Matrix models of 4-dimensional quantum Hall fluids},\hspace{3pt}  hep-th/0209182, September 2002.
\bibitem{Ch3}Y.X. Chen, \hspace{3pt} \emph{Quasi-particle excitations and hierarchies of 4-dimensional quantum Hall fluid states in the matrix models},\hspace{3pt} 
hep-th/0210059, October 2002.
\bibitem{Ch1}Y.X. Chen, B.Y. Hou and B. Y. Hou,\hspace{3pt}  \emph{Non-commutative geometry of 4-dimensional quantum Hall droplet}, \hspace{3pt} Nuclear Physics,
\textbf{B638}, 220-242, 2002.
\bibitem{El1}H. Elvang and J. Polchinski, \hspace{3pt} \emph{The Quantum Hall Effect on $\R^4$}, \hspace{3pt} hep-th/0209104, September 2002.
\bibitem{Fa1}M. Fabinger,\hspace{3pt}\emph{Higher-Dimensional Quantum Hall Effect in String Theory}, \hspace{3pt} JHEP, \textbf{0205}, 037-049, 2002.
\bibitem{Ha1}F.D.M. Haldane, \hspace{3pt}\emph{Fractional Quantization of the Hall Effect: A Hierarchy of Incompressible Quantum Fluid States},
\hspace{3pt}Physical Review Letters, \textbf{51}, 605-608, 1983.
\bibitem{Zh1}J.-P. Hu and S.-C. Zhang, \hspace{3pt}\emph{Collective excitations at the boundary of a 4D quantum Hall
droplet},\hspace{3pt}con-mat/0112432, 2001.
\bibitem{Ka1}D. Karabali and V.P. Nair,\hspace{3pt} \emph{ Quantum Hall Effect in Higher Dimensions},\hspace{3pt}  hep-th/0203264, March 2002.
\bibitem{Ki2}Y. Kimura, \hspace{3pt} \emph{Non-commutative Gauge Theory on Fuzzy Four-Sphere and Matrix Model}, \hspace{3pt} Nuclear Physics, \textbf{B637},
177-198, 2002.
\bibitem{La1}R. Laughlin, \hspace{3pt}\emph{Anomalous Quantum Hall Effect: An Incompressible Quantum Fluid with Fractionally Charged
Excitations},\hspace{3pt} Physical Review Letters,
\textbf{50}, 1395-1398. 1983. 
\bibitem{Pr1}R. Prange and S. Girvin,\hspace{3pt} \emph{The Quantum Hall Effect}, \hspace{3pt}Springer Verlag,
Berlin, Germany, 1990.   
\bibitem{Sp1}G.A.J. Sparling,\hspace{3pt}
\emph{Twistor theory and the four-dimensional Quantum Hall Effect of Zhang and Hu}, \hspace{3pt}University of Pittsburgh preprint, March 2002.
\bibitem{St1}M. Stone,\hspace{3pt} \emph{Schur functions, chiral bosons and the quantum-Hall-effect edge states},\hspace{3pt} Physical Review,
\textbf{42B}, 8399-8404, 1990.
\bibitem{Su1}L. Susskind,\hspace{3pt}  \emph{The Quantum Hall Fluid and Non-Commutative Chern-Simons Theory}, \hspace{3pt}  hep-th/0101029, January 2001.   
\bibitem{Ya1}C.N. Yang,\hspace{3pt} \emph{Generalization of Dirac's monopole to $\mathcal{SU}_2$ gauge fields}, \hspace{3pt} Journal of
Mathematical Physics,
\textbf{19}, 320-328, 1978. 
\bibitem{Zh3}S.-C. Zhang, \hspace{3pt}\emph{The Chern-Simons-Landau-Ginzburg theory of the fractional quantum Hall
effect},\hspace{3pt}International Journal of Modern Physics,
\textbf{6B}, 25-58, 1992. 
\bibitem{Zh4}S.-C. Zhang, \hspace{3pt} \emph{Exact microscopic wave function for a topological quantum membrane},\hspace{3pt}  cond-mat/0210604, October
2002.
\bibitem{Zh5}S.-C. Zhang,\hspace{3pt} \emph{ To see a world in a grain of sand},\hspace{3pt} hep-th/0210162, October 2002.
\bibitem{Ha2}S.W. Hawking and R. Penrose, \hspace{3pt}\emph{The Nature of Space and Time}, Princeton University Press, Princeton, 2000.
\bibitem{Ne1}E.T. Newman and R. Penrose, \hspace{3pt}\emph{An approach to gravitational radiation by a method of spin coefficients}, \hspace{3pt} Journal of
Mathematical Physics, \textbf{3}, 566-578, 1962.
\bibitem{Pe1}R. Penrose and W. Rindler,\hspace{3pt} \emph{ Spinors and space-time Volume 1: Two-spinor calculus and relativistic
fields},
\hspace{3pt} Cambridge University Press, Cambridge, 1984.
\bibitem{Pe2}R. Penrose and W. Rindler,\hspace{3pt} \emph{ Spinors and space-time Volume 2: Spinor and twistor methods in space-time
geometry},
\hspace{3pt} Cambridge University Press, Cambridge, 1986.
\bibitem{Sp2}D. Kapadia  and G.A.J. Sparling,\hspace{3pt}\emph{Glitch metrics}, \hspace{3pt}Laboratory of Axiomatics preprint,
2002.
\bibitem{Sp3}D. Kapadia  and G.A.J. Sparling, \hspace{3pt}\emph{A class of conformally Einstein metrics}, \hspace{3pt}Classical and Quantum
Gravity,
\textbf{24}, 4765-4776, 2000.
\bibitem{Sp4}G.A.J. Sparling,\hspace{3pt}
\emph{Zitterbewegung}, \hspace{3pt}Seminaires et Congres, Societe Mathematique de
France, \textbf{4}, 275-303, 2000.
\bibitem{Ab1}E. Abbott and B. Hoffman\hspace{3pt} \emph{Flatland: A Romance of Many Dimensions},\hspace{3pt}  Dover, New York, 1992.
\bibitem{Di1}P.A.M. Dirac, \hspace{3pt} \emph{Quantized Singularities in the Electromagnetic Field},\hspace{3pt} Proceedings of the Royal Society of London,
\textbf{A133}, 60-72, 1931.
\bibitem{Di2}P.A.M. Dirac, \hspace{3pt} \emph{The Theory of Magnetic Poles} \hspace{3pt}Physical Review, \textbf{74}, 817-830, 1948.
\bibitem{Pe5}R. Penrose and G.A.J. Sparling, \hspace{3pt}\emph{The anti-self-dual Coulomb field's non-Hausdorff twistor space}, in \emph{Further advances in
twistor theory, Volume I: The Penrose transform and its applications}, editors L.J. Mason and L.P. Hughston, Longman, Harlow, 1990.
\bibitem{Ma1}L.J. Mason, \hspace{3pt}\emph{Twistor blisters}, in \emph{Further advances in
twistor theory, Volume I: The Penrose transform and its applications}, editors L.J. Mason and L.P. Hughston, Longman, Harlow, 1990.
\bibitem{Ca1}E. Cartan, \hspace{3pt} \emph{Le principe de dualite et la theorie des groupes simple et semi-simples},\hspace{3pt} Bulletin de  Science Mathematique,
\textbf{49}, 361-374, 1925.
\bibitem{Gu1}M. Gunaydin,\hspace{3pt}  \emph{Generalized conformal and superconformal group actions and Jordan algebras},\hspace{3pt} Modern Physics Letters,
\textbf{A8}, 1407-1416, 1993.
\bibitem{Gu2}M. Gunaydin, K. Koepsell and H. Nicolai,\hspace{3pt}   \emph{Conformal and quasiconformal realizations of exceptional Lie groups},
\hspace{3pt}hep-th/0008063, 2000.
\bibitem{Gu3}M. Gunaydin, K. Koepsell and H. Nicolai,\hspace{3pt}   \emph{The minimal unitary representation of $E_{8(8)}$},\hspace{3pt} Advances in Theoretical
and Mathematical Physics, \textbf{5}, 923-946, 2002.
\bibitem{Ba1}J. Baez,\hspace{3pt} \emph{The Octonions},\hspace{3pt}Bulletin of the American Mathematical Society, \textbf{39}, 145-205, 2002.
\bibitem{Ma2}C. Manogue and T. Dray, \hspace{3pt} \emph{Octonionic Mobius transformations},\hspace{3pt} Modern Physics Letters, \textbf{A14}, 1243-1256, 1999.
\bibitem{Fr1}H. Freudenthal, \emph{Ausnahmegruppen und Oktavengeometrie}, \hspace{3pt}Geometriae Dedicata, \textbf{19}, 7-63, 1985.
\bibitem{Hu1}L.P. Hughston, \emph{Twistors and Particles}, \hspace{3pt}Lecture Notes in Physics \textbf{97}, Springer-Verlag, Berlin 1979.
\bibitem{Sp6}Z. Perjes and G.A.J. Sparling, \hspace{3pt}\emph{The twistor structure of hadrons, KFKI-76-62},  \hspace{3pt}in  \hspace{3pt}\emph{Advances in Twistor
Theory}, \hspace{3pt}editors L.P. Hughston and R.S. Ward, Pitman Press, London, 1979.
\bibitem{Sp5}G.A.J. Sparling, \hspace{3pt} \emph{Theory of massive particles. {I}. {A}lgebraic structure}, \hspace{3pt}Philosophical
Transactions of the Royal Society of London,
\textbf{A301}(1458), 27-74, 1981.
\bibitem{Sm1}T. Smith, \hspace{3pt}\emph{WWW Home page}, \hspace{3pt}www.innerx.net/personal/tsmith/TShome.html.
\bibitem{Jo1}P. Jordan, J. von Neumann, E. P. Wigner, \hspace{3pt}\emph{On an Algebraic Generalization of the Quantum Mechanical Formalism}, \hspace{3pt}Annals
of Mathematics, \textbf{35(1)}, 29-64, 1934.
\bibitem{Wi1}E.P. Wigner, \hspace{3pt}\emph{On Unitary Representations of the Inhomogeneous Lorentz Group}, \hspace{3pt}Annals of Mathematics, \textbf{40(1)},
149-204, 1939.
\bibitem{Wo2}N.M.J. Woodhouse,\hspace{3pt} \emph{Geometric Quantization, Second edition},\hspace{3pt} Clarendon Press, Oxford, 1997.

\end{thebibliography}
\end{document}